\documentclass[aps,twocolumn,showpacs,superscriptaddress]{revtex4-1}

\usepackage{bm,graphicx,textcomp,amssymb,amsmath,dcolumn,xcolor}
\graphicspath{{figs/}}

\newcommand{\ket}[1]{\left|#1\right>}
\newcommand{\bra}[1]{\left<#1\right|}

\newcommand{\nn}{\nonumber\\}

\newcommand{\f}[1]{\mbox{\boldmath$#1$}}

\newcommand{\bea}{\begin{eqnarray}}
\newcommand{\ea}{\end{eqnarray}}
\newcommand{\eea}{\end{eqnarray}}
\newcommand{\ord}{\,{\cal O}}

\begin{document}

\title{Boltzmann relaxation dynamics of strongly interacting spinless fermions on a lattice}

\author{Friedemann Queisser}

\affiliation{Helmholtz-Zentrum Dresden-Rossendorf, 
  Bautzner Landstra{\ss}e 400, 01328 Dresden, Germany,}

\affiliation{Institut f\"ur Theoretische Physik, 
  Technische Universit\"at Dresden, 01062 Dresden, Germany.}

\affiliation{Fakult\"at f\"ur Physik,
  Universit\"at Duisburg-Essen, Lotharstra{\ss}e 1, 47057 Duisburg, Germany,}

\author{Sebastian Schreiber}

\affiliation{Fakult\"at f\"ur Physik,
  Universit\"at Duisburg-Essen, Lotharstra{\ss}e 1, 47057 Duisburg, Germany,}

\author{Peter Kratzer}

\affiliation{Fakult\"at f\"ur Physik,
  Universit\"at Duisburg-Essen, Lotharstra{\ss}e 1, 47057 Duisburg, Germany,}

\author{Ralf Sch\"utzhold}

\affiliation{Helmholtz-Zentrum Dresden-Rossendorf, 
  Bautzner Landstra{\ss}e 400, 01328 Dresden, Germany,}

\affiliation{Institut f\"ur Theoretische Physik, 
  Technische Universit\"at Dresden, 01062 Dresden, Germany.}

\affiliation{Fakult\"at f\"ur Physik,
  Universit\"at Duisburg-Essen, Lotharstra{\ss}e 1, 47057 Duisburg, Germany,}

\date{\today}

\begin{abstract}
  Motivated by the recent interest in non-equilibrium phenomena in quantum
  many-body systems, we study strongly interacting fermions on a lattice by
  deriving and numerically solving quantum Boltzmann equations that describe
  their relaxation to thermodynamic equilibrium.
  The derivation is carried out by inspecting the hierarchy of correlations
  within the framework of the $1/Z$-expansion.
  Applying the Markov approximation, we obtain the dynamic equations for
  the distribution functions. 
  Interestingly, we find that in the strong-coupling limit, collisions between particles and holes dominate over particle-particle and hole-hole collisions -- 
  in stark contrast to weakly interacting systems. 
  As a consequence, our numerical simulations show that the relaxation time scales strongly depend on the type of excitations (particles or holes or both) that are initially present. 

\end{abstract}


\maketitle

\section{Introduction}

In interacting quantum many-body systems, the nature of the excitations and their 
relaxation to the thermodynamic equilibrium state {\color{black}\cite{L1876,B1877}} display a large diversity, 
and may vary from system to system, depending both on the dimensionality and on 
the specific interactions {\color{black} \cite{D91,S94,RDO02,CR10,RS12,R13,PSSV11,RDO08,Getal12,Ketal16,Netal16,KWW06}}. 
Here, we are interested in quantum systems in which the  excitations can be described 
as quasiparticles. 
This raises the question of whether the quasiparticles interact in such a way that 
their incoherent scattering processes finally lead to their equilibration. 
In other words, we ask if, and under which conditions, it is possible to construct 
a quantum Boltzmann equation that describes the collisions between quasiparticles.
For weak interactions, a famous example is of course the Fermi liquid \cite{Landau57,Landau59} 
of electrons subject to Coulomb interaction in three spatial dimensions. 
The quasiparticles are electrons and holes, possibly with a renormalized mass, 
that interact via two-particle scattering in a screened Coulomb potential \cite{Pines66}. 
However, for strongly interacting electrons, as typically found in transition-metal 
oxides or nitrides, already the electronic ground state may differ strongly 
from the weakly interacting case
(for a review, see e.g. \cite{IFT98}), and the relaxation kinetics of quasiparticle 
excitations remains elusive. 
Hence, the general principles of the non-equilibrium relaxation are an active research topic up to date \cite{Kemper2018}.
In the following, we restrict our discussion to closed quantum lattice systems 
without disorder and dissipation. 
This means that equilibration is supposed to proceed solely by intrinsic interactions. 
Yet, there is rich physics to be found: 
While the lifetime of excitations in a Fermi liquid follows a generic law, 
it turns out that relaxation in a quantum system with strong interactions 
may proceed via several intermediate stages and thus on widely different time scales, 
see, e.g., \cite{EKW09,EKW10,WEFWBH18,PSP18,GBEW19}.

In this work, we show for a particular example that even in the strongly interacting 
limit the kinetic equation describing thermalization still has the mathematical 
structure of a quantum Boltzmann equation, albeit with a different physical 
interpretation of the collision term,  
see also \cite{B75}. 
Specifically, we study a lattice model of spinless Fermions with interactions 
between neighboring lattice sites. 
In the limit of strong interactions, giving rise to a gapped excitation spectrum, 
we find that electron-hole scattering is the dominant relaxation mechanism, 
in striking contrast to the conventional Fermi liquid, where hole-hole and 
electron-electron interactions contribute on equal footing with electron-hole 
interactions to the overall relaxation rate. 

Spinless Fermions are considered as a very simple model epitomizing 
the features of a metal-insulator transition \cite{WCT14,NNOF18}. 
In applications to the electronic structure of materials, 
the model may be applicable to crystalline solids with partial band filling in the independent electron approximation, but 
strong on-site Coulomb repulsion,  
which guarantees that each lattice site 
will be occupied only once, and double occupancy by electrons of opposite 
spin can be ignored at sufficiently low excitation energies. 
In addition, in a solid with a less than 
half-filled band, sufficiently  strong on-site Coulomb interaction gives rise to a ferromagnetic ground state \cite{Vollhardt97}; 
i.e., all spins are aligned and, as a first approximation, 
the spin degree of freedom can be neglected. 
It is noteworthy that ultra-cold atoms in an optical lattice offer another 
possibility to realize the model studied here, provided that the repulsion 
between atoms at the same lattice site is sufficiently strong to preclude 
multiple occupation. 
In this case, the trapped atoms may be considered {\em effectively} 
as spinless Fermions independent of their actual spin. 
In addition, we note that, in the context of ultra-cold atoms, spinless fermions 
  interacting indirectly via additional bosons have been considered~\cite{Polak,Buechler}. 

In the center of our interest are lattice systems with a high coordination 
number $Z$ (which means in practice, lattices in high dimensions). 
This is in contrast to the physics in one-dimensional systems, 
where the quasiparticle picture is often not suitable as a starting 
point for further analysis. 
The peculiar thermalization behavior of one-dimensional systems 
has been extensively studied in recent work,  
see, e.g., 
\cite{MWNM98,R09a,R09b,RDYO07,BKL10,KISD11,SVPH14}.
\textcolor{black}{In the opposite limit of high dimensionality, several authors used non-equilibrium Green functions 
  on the two-time Keldysh contour as the starting point of their description. The Kadanoff-Baym equation \cite{Kadanoff-Baym} for these Green functions 
  is then solved either numerically or with the help of some approximations, see e.g. Ref.~\onlinecite{Bonitz}. In popular approaches, the thermalization behavior 
  is described by a self energy that is taken from dynamic mean field theory \cite{Aoki2014}. While this is appropriate in the limit of very high dimensionality, it usually neglects momentum conservation and the dependence of the scattering rates, and hence the self energy, on the momentum 
  transfer in the collision~\cite{WEFWBH18}.
  Finally, schemes based on the usual BBGKY hierarchy can only treat weak and
  moderate interactions, see, e.g., Ref.~\onlinecite{B75}, while we are mainly interested 
  in strong interactions. 
  Thus, derivations of Boltzmann equations for strongly interacting many-body 
  systems on higher-dimensional lattices, as considered here, would be highly non-trivial 
  using other methods.}

The structure of the paper is as follows: 
After defining the model of spinless Fermions, we briefly recapitulate 
the derivation of the Boltzmann equation \cite{B75} in the weakly interacting 
case, making use of the Born-Markov approximation \cite{BP02,RK02}. 
Next, we introduce correlation functions in the spirit of the BBGKY hierarchy 
of non-equilibrium statistical mechanics \cite{K46,B46,BG46},  
but with the important difference that we consider 
the correlations between lattice sites instead of those between particles. 
We show that an expansion in $1/Z$ allows us to define the spectrum of 
quasiparticle excitations at the $Z^{-1}$ level, while the higher orders 
of the expansion give rise to interactions among quasiparticles 
and offer a natural way to close the BBGKY-like hierarchy of equations. 
We discuss the solutions for translationally invariant systems, 
in particular for the non-trivial example of the correlated ground 
state on a bipartite lattice. 
Kinetic equations are worked out explicitly for the limit of strong interaction. 
If, in addition, the interaction is also short-ranged and extends to nearest-neighbor 
lattice sites only, the interaction between quasiparticles is found 
to be strongly anisotropic. 
We illustrate the consequences of the anisotropic interactions by 
numerically solving the kinetic equation. 
In contrast to the well-known Fermi liquid with isotropic Coulomb interaction, 
thermalization of quasiparticles 
displays several time scales due to the dependence of the collisions 
on momentum transfer. 
In particular, this behavior is observed when the initial distributions 
of quasiparticles and quasiholes differ strongly from each other.

\section{The Model}

We consider spinless Fermions \cite{IFT98,WCT14} moving on a lattice given by the 
hopping matrix $J_{\mu\nu}$ and repelling each other via the Coulomb 
matrix $V_{\mu\nu}$
\begin{align}
  \label{spinlessfermions}
  \hat{H}
  =
  -\frac{1}{Z}\sum_{\mu,\nu}J_{\mu\nu}\hat c^\dagger_\mu \hat c_\nu
  +\frac{1}{2Z}\sum_{\mu,\nu}V_{\mu\nu}\hat n_\mu \hat n_\nu 
  \,.
\end{align}
As usual, $\hat c^\dagger_\mu$ and $\hat c_\nu$ are the fermionic creation 
and annihilation operators for the lattice sites $\mu$ and $\nu$ with the
corresponding number operators $\hat n_\mu=\hat c^\dagger_\mu\hat c_\mu$.
Furthermore, $Z$ denotes the coordination number of the translationally 
invariant lattice, i.e., the number of nearest neighbors. 
In the following, we consider nearest-neighbor interaction and tunneling 
for simplicity, but our results can be generalized in a straight forward 
manner.  

In the limit of small interactions $V_{\mu\nu}$, the ground state 
of~\eqref{spinlessfermions} can be described by a Fermi gas 
and is thus metallic for $0<\langle\hat n_\mu\rangle<1$. 
For large interactions $V_{\mu\nu}$, however, the structure of the ground 
state changes.
Assuming half filling and a bipartite lattice, we have a spontaneous 
breaking of the translational symmetry where one sub-lattice is occupied 
while the other sub-lattice is empty 
(up to small virtual tunneling corrections), which is usually referred to 
as a charge density wave -- quite analogous to the famous Mott insulator 
state in the Fermi-Hubbard model, see, e.g., \cite{H63}. 

\section{Weak interaction limit}

Let us start by briefly recapitulating the conventional derivation of the 
Boltzmann equation in the limit of weak interactions, see, e.g., \cite{RK02}.
After a spatial Fourier transform $\hat c_\mu\to\hat c_\mathbf{k}$, 
the relevant distribution functions $f_\mathbf{k}$ are just the 
occupation numbers per mode $\mathbf{k}$ and their time derivative reads 
according to \eqref{spinlessfermions}
\bea
\label{distributionfunction}
&&i\partial_t f_\mathbf{k}
=
i\partial_t\langle \hat c_\mathbf{k}^\dagger \hat c_\mathbf{k}\rangle
=
-\int_{\mathbf{p}}\int_{\mathbf{q}}
V_\mathbf{q}\times
\nn
&&
\left(\langle
  \hat c^\dagger_\mathbf{k}\hat c^\dagger_\mathbf{p}
  \hat c_\mathbf{p+q}\hat c_\mathbf{k-q}
  \rangle^\mathrm{corr}
  -
  \langle
  \hat c^\dagger_\mathbf{k-q}\hat c^\dagger_\mathbf{p+q} 
  \hat c_\mathbf{p}\hat c_\mathbf{k}
  \rangle^\mathrm{corr}
\right) 
\,,
\ea
where we have defined the four-momentum correlators via 
$\langle
\hat c^\dagger_\mathbf{k}\hat c^\dagger_\mathbf{p}\hat c_\mathbf{k'}\hat c_\mathbf{p'}
\rangle^\mathrm{corr}
=
\langle
\hat c^\dagger_\mathbf{k}\hat c^\dagger_\mathbf{p}\hat c_\mathbf{k'}\hat c_\mathbf{p'}
\rangle
+
\langle\hat c^\dagger_\mathbf{k}\hat c_\mathbf{k'}\rangle
\langle\hat c^\dagger_\mathbf{p}\hat c_\mathbf{p'}\rangle
-
\langle\hat c^\dagger_\mathbf{k}\hat c_\mathbf{p'}\rangle
\langle\hat c^\dagger_\mathbf{p}\hat c_\mathbf{k'}\rangle$.
To first order in the interaction strength $V_\mathbf{q}$, 
their time derivative reads
\bea
\label{fourpointpert}
&&i\partial_t \langle
\hat c^\dagger_{\mathbf{k}}c^\dagger_{\mathbf{p}} 
c_{\mathbf{p+q}}c_{\mathbf{k-q}}
\rangle^\mathrm{corr}
=
\nn
&&(J_\mathbf{k}+J_\mathbf{p}-J_\mathbf{k-q}-J_\mathbf{p+q})
\langle
\hat c^\dagger_{\mathbf{k}}c^\dagger_{\mathbf{p}} 
c_{\mathbf{p+q}}c_{\mathbf{k-q}}
\rangle^\mathrm{corr}
-
\nonumber\\
&&
(V_\mathbf{q}-V_\mathbf{k-p-q})\times 
\nn
&&\left[f_\mathbf{k}f_\mathbf{p}(1-f_\mathbf{k-q})(1-f_\mathbf{p+q})
  -
  (f_\mathbf{k}f_\mathbf{p}\leftrightarrow f_\mathbf{k-q}f_\mathbf{p+q})
\right].
\ea
Abbreviating these four-momentum correlators by 
$C_{\mathbf{k}\mathbf{p}\mathbf{q}}$,
the above equation can be cast into the simple form 
$i\partial_tC_{\mathbf{k}\mathbf{p}\mathbf{q}}=
\Omega_{\mathbf{k}\mathbf{p}\mathbf{q}}C_{\mathbf{k}\mathbf{p}\mathbf{q}}
-S_{\mathbf{k}\mathbf{p}\mathbf{q}}$
with the source term $S_{\mathbf{k}\mathbf{p}\mathbf{q}}$ 
containing the distribution functions $f_\mathbf{k}$. 
Formally, this linear equation has the retarded solution
\bea
\label{retarded}
C_{\mathbf{k}\mathbf{p}\mathbf{q}}(t)
=
i\int\limits_{-\infty}^t dt'\,S_{\mathbf{k}\mathbf{p}\mathbf{q}}(t')
\exp\left\{
  -i\Omega_{\mathbf{k}\mathbf{p}\mathbf{q}}(t-t')
\right\}
\,.
\ea
In order to arrive at the Boltzmann equation which is local in time, 
we now employ the Markov approximation 
$S_{\mathbf{k}\mathbf{p}\mathbf{q}}(t')\approx S_{\mathbf{k}\mathbf{p}\mathbf{q}}(t)$
in the above integrand, which can be motivated by the fact that the 
distribution functions are slowly varying.   
Then \eqref{retarded} can be solved approximately 
\bea
\label{approximately}
C_{\mathbf{k}\mathbf{p}\mathbf{q}}(t)
\approx
\frac{S_{\mathbf{k}\mathbf{p}\mathbf{q}}(t)}
{\Omega_{\mathbf{k}\mathbf{p}\mathbf{q}}-i\varepsilon}
\,,
\ea
where the infinitesimal convergence factor $\varepsilon>0$ is inserted 
in order to pick out the retarded solution.
As usual, the limit $\varepsilon\downarrow0$ yields the principal value 
plus a delta distribution.
The principal value corresponds to the adiabatic solution of 
$i\partial_tC_{\mathbf{k}\mathbf{p}\mathbf{q}}=
\Omega_{\mathbf{k}\mathbf{p}\mathbf{q}}C_{\mathbf{k}\mathbf{p}\mathbf{q}}
-S_{\mathbf{k}\mathbf{p}\mathbf{q}}\approx0$, 
while the delta distribution contributes at 
$\Omega_{\mathbf{k}\mathbf{p}\mathbf{q}}=0$ where adiabaticity breaks down.
This is the term which generates the Boltzmann collision term, 
where $\Omega_{\mathbf{k}\mathbf{p}\mathbf{q}}=0$ corresponds to energy 
conservation. 
Inserting the approximate solution \eqref{approximately} 
back into \eqref{distributionfunction} 
yields the well-known Boltzmann equation (see, e.g., \cite{RK02}) 
\bea
\label{smallV}
&&
\partial_t f_\mathbf{k}
=
-2\pi\int_{\mathbf{p}}\int_{\mathbf{q}}
V_\mathbf{q}(V_\mathbf{q}-V_\mathbf{k-p-q})
\times
\nn
&&
\delta(J_\mathbf{k}+J_\mathbf{p}-J_\mathbf{k-q}-J_\mathbf{p+q})
\times
\nonumber\\
&&
\left[f_\mathbf{k}f_\mathbf{p}(1-f_\mathbf{k-q})(1-f_\mathbf{p+q})
  -
  (f_\mathbf{k}f_\mathbf{p}\leftrightarrow f_\mathbf{k-q}f_\mathbf{p+q})
\right].
\ea
Here $\mathbf{q}$ denotes the momentum transfer, i.e., particles with 
initial momenta $\mathbf{k}$ and $\mathbf{p}$ collide and are scattered 
to the final momenta $\mathbf{k-q}$ and $\mathbf{p+q}$ or vice versa.
The delta distribution in the second line represents energy conservation 
in such a collision process and the factor in the first line 
yields the differential cross section. 
As is well known, this equation respects the conservations laws 
of energy, momentum and probability, as well as consistency conditions 
(such as the crossing relation) and has far reaching consequences 
such as the $H$-theorem, see, e.g., \cite{B1872}.

\section{Hierarchy of Correlations}

In the above derivation, we exploited the assumption of weak interaction in 
two ways: first, by employing a perturbative expansion in $V_{\mu\nu}$ in 
equation~\eqref{fourpointpert}, and, second, by applying the Markov 
approximation~\eqref{approximately}.
This approximation is based on the separation of time scales, i.e., 
the distribution functions $f_\mathbf{k}$ are slowly varying 
(on a time scale set by $V_{\mu\nu}$) in comparison to the rapid 
oscillations in~\eqref{retarded} with the frequencies 
$\Omega_{\mathbf{k}\mathbf{p}\mathbf{q}}$ which are set by $J_{\mu\nu}$. 
For strong interactions, this procedure is no longer applicable. 
However, we will show in the following that the coordination number 
of lattice sites in high dimensions can 
be used in a similar way to establish a systematic expansion. 

The framework for deriving this expansion is provided by  
the hierarchy of correlations 
\cite{NS10,QNS12,QKNS14,KNQS14,NQS14,NQS16,environment,arxiv}. 
In this approach one  considers the reduced 
density matrices $\hat\rho_\mu$ for one site, $\hat\rho_{\mu\nu}$ for two 
sites, and so on.  
Multi-site density matrices will in general not be simple products of the single-site quantities. 
We therefore split off the difference due to correlations between sites, i.e. we write
$\hat\rho_{\mu\nu}^{\rm corr}=\hat\rho_{\mu\nu}-\hat\rho_\mu\hat\rho_\nu$,  
and analogously for multi-site correlations. 
The time-dependence of  correlations can be cast into the following hierarchy of evolution equations 
\bea
\label{on-site}
\partial_t\hat\rho_\mu
&=&
f_1(\hat\rho_\nu,\hat\rho_{\mu\nu}^{\rm corr})
\,,
\\
\label{two-site}
\partial_t\hat\rho_{\mu\nu}^{\rm corr}
&=&
f_2(
\hat\rho_\nu,\hat\rho_{\mu\nu}^{\rm corr},\hat\rho_{\mu\nu\sigma}^{\rm corr})
\,,
\\
\label{three-site}
\partial_t\hat\rho_{\mu\nu\sigma}^{\rm corr}
&=&
f_3(
\hat\rho_\nu,\hat\rho_{\mu\nu}^{\rm corr},\hat\rho_{\mu\nu\sigma}^{\rm corr},
\hat\rho_{\mu\nu\sigma\lambda}^{\rm corr})
\,,
\\
\label{four-site}
\partial_t\hat\rho_{\mu\nu\sigma\lambda}^{\rm corr}
&=&
f_4(
\hat\rho_\nu,\hat\rho_{\mu\nu}^{\rm corr},\hat\rho_{\mu\nu\sigma}^{\rm corr},
\hat\rho_{\mu\nu\sigma\lambda}^{\rm corr}, 
\hat\rho_{\mu\nu\sigma\lambda\zeta}^{\rm corr})
\,,
\ea
and in complete analogy for the higher correlators \cite{NS10}. 

To derive a systematic expansion, we consider the 
hierarchy of correlations in the formal limit of large coordination numbers 
$Z\to\infty$.
Following ref.~\onlinecite{NS10}, it can be shown that the $n$-site 
correlators are by a factor $1/Z$ smaller than $n-1$-site correlators. 
For instance, starting from the on-site density matrix $\hat\rho_{\mu}=\ord(Z^0)$ 
as the zeroth order, 
two-site correlators are smaller, $\hat\rho_{\mu\nu}^{\rm corr}=\ord(1/Z)$. 
Furthermore, the three-site correlators are suppressed even stronger via 
$\hat\rho_{\mu\nu\sigma}^{\rm corr}=\ord(1/Z^2)$, and so on.
The decreasing role of higher-order correlators justifies an approximative 
scheme based on a truncation of the hierarchy at some specific level even 
without having to invoke any separation of time scales, ergodicity, 
or other supportive arguments. 
In physics language, an iterative 
approximation scheme can be described as follows: 
We start from a mean-field solution $\hat\rho_\mu^0$ which is obtained 
to zeroth order in $1/Z$ by neglecting $\hat\rho_{\mu\nu}^{\rm corr}$ 
on the right-hand side of~\eqref{on-site} and equating 
$\partial_t\hat\rho_\mu\approx f_1(\hat\rho_\nu,0)$. 
Next, we insert this solution $\hat\rho_\mu^0$ 
into~\eqref{two-site} and obtain, to first order in $1/Z$, the 
approximation 
$\partial_t\hat\rho_{\mu\nu}^{\rm corr}\approx
f_2(\hat\rho_\nu^0,\hat\rho_{\mu\nu}^{\rm corr},0)$
This provides us with a set of inhomogeneous linear differential equations for the 
two-point correlations $\hat\rho_{\mu\nu}^{\rm corr}$.
The stationary solutions of this set can be considered as the quasi-particle modes; and 
in this way the quasi-particle energy spectrum is obtained. 

However, aiming at the derivation of a quantum Boltzmann equation, 
it is clear that we have to go further. 
This can be understood from the following considerations: 
The quasi-particles resulting from a truncation at the level 
of~\eqref{two-site} are non-interacting; hence 
~\eqref{two-site} is insufficient to derive a Boltzmann collision term to first order in $1/Z$. 
In other words, 
a set of differential equations linear in the variable 
$\hat\rho_{\mu\nu}^{\rm corr}$, such as $\partial_t\hat\rho_{\mu\nu}^{\rm corr}\approx
f_2(\hat\rho_\nu^0,\hat\rho_{\mu\nu}^{\rm corr},0)$ can not describe 
collisions. 
Therefore, we need to study higher orders in $1/Z$ and and 
interpret the interactions between the quasi-particles arising 
on this level as collision terms. 
The above derivation of a quantum Boltzmann equation for weak interactions already suggests that 
one should not stop on the level of the three-point correlators $\hat\rho_{\mu\nu\sigma}^{\rm corr}$ 
that enter into the right-hand side of~\eqref{two-site}. 
Due to the structure of the Coulomb interactions,  
we have to include the four-point correlators, too, in order to derive the 
Boltzmann equation (see below). 

To arrive at a consistent treatment up to order $1/Z^2$, 
one should also insert the solution for $\hat\rho_{\mu\nu}^{\rm corr}$, 
once it has been obtained, back into equation~\eqref{on-site}. 
A similar argument can be applied to $\hat\rho_{\mu\nu\sigma}^{\rm corr}$ 
which should be inserted into 
\eqref{two-site} to obtain an improved quasi-particle spectrum. 
We speak of this procedure as taking into account the back-reactions. 
In physical terms, this amounts to a renormalization of the mean-field 
description by $\hat\rho_\mu^0$ due to the 
quasi-particle fluctuations. 
For the case considered in the following application, 
a small perturbation around the charge-density wave state at half filling, 
the back-reactions play a minor role, and they will be omitted in the following. 

\subsection{Translation-invariant systems}

In this subsection, we specialize to spatially homogeneous systems. 
Consequently, we re-formulate the equations in the Fourier space of wave vectors.  
As starting point of the hierarchy, we first need to specify  the on-site 
density matrix $\hat\rho_\mu$ or its zeroth-order (mean-field) approximation 
$\hat\rho_\mu^0$.
Further specializing to the case of a half-filled band, equation~\eqref{on-site} 
has the simple solution 
\bea
\label{homogeneous-background}
\hat\rho_\mu=\frac12\left(\ket{0}\bra{0}+\ket{1}\bra{1}\right)
=\frac12\,\f{1}_\mu
\,.
\ea
Due to the assumed spatial homogeneity and particle number conservation, 
this solution is actually unique and hence there is no back-reaction; i.e.  
$\hat\rho_\mu=\hat\rho_\mu^0$.

As in Eq.~\eqref{distributionfunction}, 
the distribution functions $f_\mathbf{k}$ are given by the relevant 
two-point correlations 
$\langle \hat c_{\mu}^\dagger \hat c_{\nu}\rangle^\mathrm{corr}$
via 
\bea
\label{homogeneous-two-point}
\langle \hat c_{\mu}^\dagger \hat c_{\nu}\rangle=
\langle \hat c_{\mu}^\dagger \hat c_{\nu}\rangle^\mathrm{corr}
+\delta_{\mu\nu}\langle \hat{n}_\mu\rangle
=
\int_\mathbf{k} 
f_\mathbf{k}e^{i\mathbf{k}\cdot(\mathbf{x}_{\nu}-\mathbf{x}_{\mu})}
\,.
\ea
Then, equation~\eqref{two-site} implies 
\bea
\label{twopointfourier}
i\partial_t f_\mathbf{k}
=
\int_\mathbf{q} 
V_{\mathbf{k}+\mathbf{q}}
(g_{\mathbf{q}\mathbf{k}}-g_{\mathbf{k}\mathbf{q}})
\,,
\ea
where the $g_{\mathbf{q}\mathbf{k}}$ denote the Fourier components of the 
relevant three-point correlations 
\bea
\label{homogeneous-three-point}
\langle\hat n_\alpha\hat c_{\beta}^\dagger \hat c_{\gamma}\rangle^\mathrm{corr}
=
\int_\mathbf{p}\int_\mathbf{q} 
g_{\mathbf{p}\mathbf{q}}
e^{i\mathbf{p}\cdot(\mathbf{x}_{\beta}-\mathbf{x}_{\alpha})
  +
  i\mathbf{q}\cdot(\mathbf{x}_{\gamma}-\mathbf{x}_{\alpha})}
\,.
\ea
The time-derivatives of the three-point correlators can be obtained 
from~\eqref{three-site} and have a form similar to~\eqref{fourpointpert}
\bea
\label{threepointfourier}
i\partial_t g_{\mathbf{q}\mathbf{k}}
=
(J_{\mathbf{q}}-J_{\mathbf{k}})
g_{\mathbf{q}\mathbf{k}}
+S^{(3)}_{\mathbf{q}\mathbf{k}}
\,,
\ea
but, in contrast to~\eqref{fourpointpert}, the source term $S^{(3)}_{\mathbf{q},\mathbf{k}}$ 
contains four-point correlations (instead of distribution functions)  
\begin{align}
  \label{homogeneous-four-point}
  \langle
  \hat c^\dagger_{\alpha}\hat c_{\beta}\hat c^\dagger_{\mu}\hat c_{\nu}
  \rangle^\mathrm{corr}
  &=
    \int_\mathbf{k}\int_\mathbf{p}\int_\mathbf{q}
    h_{\mathbf{k}\mathbf{p}\mathbf{q}}
    \times
    \nn
  &
    e^{i\mathbf{k}\cdot(\mathbf{x}_{\alpha}-\mathbf{x}_{\nu})+
    i\mathbf{p}\cdot(\mathbf{x}_{\beta}-\mathbf{x}_{\nu})
    +i\mathbf{q}\cdot(\mathbf{x}_{\mu}-\mathbf{x}_{\nu})}
    \,.
\end{align}
Finally, their time-derivative reads according to~\eqref{four-site}
\begin{align}
  \label{fourpointFourier}
  i\partial_t 
  h_{\mathbf{k}\mathbf{p}\mathbf{q}}
  =
  (J_{\mathbf{k}}-J_{\mathbf{p}}+J_{\mathbf{q}}-J_{\mathbf{k}+\mathbf{p}+\mathbf{q}})
  h_{\mathbf{k}\mathbf{p}\mathbf{q}}
  +S^{(4)}_{\mathbf{k}\mathbf{p}\mathbf{q}}
  \,,
\end{align}
where the source term $S^{(4)}_{\mathbf{k}\mathbf{p}\mathbf{q}}$ contains 
products of two-point correlators, 
somewhat similar to~\eqref{fourpointpert}. 

Now we may integrate the evolution equations~\eqref{threepointfourier} 
and \eqref{fourpointFourier} in the same way as in~\eqref{retarded}, 
which yields a double time integral. 
In order to approximate this integral, we again use the Markov approximation:
Since the two-point correlations 
$\langle \hat c_{\mu}^\dagger \hat c_{\nu}\rangle^\mathrm{corr}$
scale with $1/Z$ but the three-point correlators 
$\langle\hat n_\alpha\hat c_{\beta}^\dagger \hat c_{\gamma}\rangle^\mathrm{corr}$
scale with $1/Z^2$, the distribution functions $f_\mathbf{k}$ are slowly varying 
according to~\eqref{twopointfourier}, because the right-hand side is suppressed 
by an additional factor of $1/Z$.
(To first order in $1/Z$, the distribution functions $f_\mathbf{k}$ are constant.) 
In contrast, the Fourier components of the three-point $g_{\mathbf{q}\mathbf{k}}$ 
and four-point $h_{\mathbf{k}\mathbf{p}\mathbf{q}}$ contributions are 
rapidly oscillating with the frequencies 
$\Omega_{\mathbf{q}\mathbf{k}}=J_{\mathbf{q}}-J_{\mathbf{k}}$
as well as 
$\Omega_{\mathbf{k}\mathbf{p}\mathbf{q}}=
J_{\mathbf{k}}-J_{\mathbf{p}}+J_{\mathbf{q}}-J_{\mathbf{k}+\mathbf{p}+\mathbf{q}}$ 
according to Eq.s~\eqref{threepointfourier} and \eqref{fourpointFourier}.
Using this separation of time scales, the double time integral can be 
evaluated within Markov approximation in analogy to~\eqref{approximately}
by simplifying the integrand according to 
$f_\mathbf{k}(t)\approx f_\mathbf{k}(t')$. 

Inserting this solution of the double time integral back into 
Eq.~\eqref{twopointfourier}, we obtain a Boltzmann equation which has exactly 
the same form as in~\eqref{smallV}. 
This is perhaps not too surprising since we did not assume that the 
interactions $V_{\mu\nu}$ are strong.
In fact, the on-site state~\eqref{homogeneous-background} could represent 
free (or weakly interacting) fermions in their ground state 
(or in a thermal state). 
As a crucial difference, however, the above derivation of the Boltzmann 
equation is based on an expansion into powers of $1/Z$ instead of 
$V_{\mu\nu}$. 
Thus, the above $1/Z$-derivation can also be applied to the strongly 
interacting case. 

\subsection{Mean-field solutions with broken symmetry}
Let us now consider the limit of strong interactions $V_{\mu\nu}$. 
\textcolor{black}{Next, we choose as a reference a mean-field solution that necessarily depends on the type of lattice and the filling factor.}
Assuming a bipartite lattice at half filling, the ground state is a
Mott-type insulator \cite{IFT98} since the fermions mainly occupy one sub-lattice,  
and tunneling to the other sub-lattice is suppressed by the repulsion 
$V_{\mu\nu}$. 
Thus, we start with the mean-field ansatz 
\bea
\label{inhomogeneous}
\hat\varrho^0_\mu=
\left\{
  \begin{array}{lll}
    \ket{0}\bra{0}=\f{1}_\mu-\hat n_\mu & {\rm for} & \mu\in{\mathcal A}
    \\
    \ket{1}\bra{1}=\hat n_\mu & {\rm for} & \mu\in{\mathcal B}
  \end{array}
\right.
\,,
\ea
where $\mathcal A$ and $\mathcal B$ denote the two sub-lattices. 
This ansatz asserts different occupation of each sub-lattice and 
thus breaks the translational symmetry of the original problem. 
Physically, this would correspond to a charge density wave.
In a square lattice or in the two-dimensional principal lattice 
planes of cubic or hyper-cubic lattices, for example, 
the fermions would form a checker-board pattern, see Fig.~\ref{figure-schach}. 

In this case, the proper treatment of the correlations 
$\langle\hat c_{\mu}^\dagger\hat c_{\nu}\rangle^\mathrm{corr}$ 
requires a case distinction.  One needs to distinguish 
which of the two sub-lattices $\mu$ and $\nu$ belong to. 
In the following, we denote these sub-lattices by calligraphic superscripts, 
e.g., for $\mu\in\mathcal A$ and $\nu\in\mathcal B$,
the expectation value 
$\langle\hat c_{\mu}^\dagger\hat c_{\nu}\rangle$ is given by 
the Fourier transform of $f^{\mathcal{AB}}_\mathbf{k}$, 
and analogously for other combinations of superscripts. 
The on-site equation~\eqref{on-site} then determines the back-reaction 
of the correlations onto the mean field via 
$i\partial_t\langle\hat n^{\mathcal A}\rangle
=
-i\partial_t\langle\hat n^{\mathcal B}\rangle
=
\int_\mathbf{k} J_\mathbf{k}
\left(
  f^\mathcal{BA}_\mathbf{k}-f^\mathcal{AB}_\mathbf{k}
\right)$, 
but we shall not consider this small small correction  in the following. 

Since we have four functions 
$f^\mathcal{AA}_\mathbf{k}$, $f^\mathcal{AB}_\mathbf{k}$,
$f^\mathcal{BA}_\mathbf{k}$, and $f^\mathcal{BB}_\mathbf{k}$,
we denote the two sub-lattices by capital superscripts such as 
$X\in\{\mathcal{A,B}\}$.
Then Eq.~\eqref{two-site} becomes 
\bea
\label{twopoint}
i\partial_t f^{XY}_\mathbf{k}
&=&
J_\mathbf{k}(f_\mathbf{k}^{\bar X Y}-f_\mathbf{k}^{X \bar Y })
-
(V^{\bar X}-V^{\bar Y})f_\mathbf{k}^{X Y}
\nn
&&
+S_\mathbf{k}^{XY},
\ea
where $\bar X$ denotes the sub-lattice opposite to $X$,
i.e., if $X=\mathcal A$ then $\bar X=\mathcal B$ and vice versa.
Furthermore, $V^\mathcal{A}$ denotes the interaction energy 
associated to sub-lattice $\mathcal A$, i.e., 
$V^\mathcal{A}=\sum_\alpha V_{\alpha\beta}\langle\hat n_\alpha\rangle/Z$
for any $\alpha\in\mathcal A$. 
For all interactions equal, this simplifies to
$V^\mathcal{A} =V\langle\hat n_\alpha\rangle$
Again, the source terms $S_\mathbf{k}^{XY}$ also contain three-point 
correlations. 

Before continuing, let us diagonalize the above linear set of 
equations (with source terms $S_\mathbf{k}^{XY}$) because the 
$f^{XY}_\mathbf{k}$ are rapidly oscillating instead of slowly varying.
This can be achieved via a rotation in the $X$-$Y$-sub-space with 
an orthogonal $2\times2$ transformation matrix $O^a_X(\mathbf{k})$ via 
$f^{ab}_\mathbf{k}=\sum_{XY}O^a_X(\mathbf{k}) f^{XY}_\mathbf{k} O^b_Y(\mathbf{k})$, 
{\textcolor{black} see the Appendix \ref{appph}.}
In terms of the rotated functions $f^{ab}_\mathbf{k}$,
the evolution equation~\eqref{twopoint} simplifies to 
\bea
\label{twopoint-rotated}
i\partial_t f^{ab}_\mathbf{k}
=
\left(E^{b}_\mathbf{k}-E^{a}_\mathbf{k}\right)f^{ab}_\mathbf{k}
+S^{ab}_\mathbf{k}
\,,
\ea
with the quasi-particle ($a=+$) and hole ($a=-$) energies 
\bea
\label{energies}
E^\pm_\mathbf{k}
=
\frac12
\left(
  V\pm\sqrt{(V^\mathcal{A}-V^\mathcal{B})^2+4J^2_\mathbf{k}}
\right) 
\,,
\ea
where we have used $V^\mathcal{A}+V^\mathcal{B}=V$ due to 
$\langle\hat n^{\mathcal A}\rangle+\langle\hat n^{\mathcal B}\rangle=1$.
This formula with its two solutions is reminiscent of the two solutions of the Fermi-Hubbard model in the Mott insulator phase \cite{LPM69} where a lower and an upper Hubbard band are formed. In the following, we speak of a quasi-particle band and a quasi-hole band referring to the $+$ and $-$ sign in~\eqref{energies}.
Apart from the gap $V^\mathcal{A}-V^\mathcal{B}$ which is basically 
the repulsion energy, the quadratic dependence on the hopping $J^2_\mathbf{k}$
indicates that (quasi) particles and quasi-holes can only move via second-order 
tunneling processes such as co-tunneling, cf.~Fig.~\ref{figure-schach}. 

We see that the functions $f^{ab}_\mathbf{k}$ in~\eqref{twopoint-rotated}
are rapidly oscillating for $a\neq b$ but slowly varying for $a=b$.
Hence the latter two are the quasi-particle ($a=b=+$) and quasi-hole 
($a=b=-$) distribution functions, which we denote by 
$f^{+}_\mathbf{k}$ and $f^{-}_\mathbf{k}$, respectively. 
Their dynamics can be derived in complete analogy to the previous case, 
cf.~Eqs.~\eqref{twopointfourier}-\eqref{fourpointFourier}, 
the only differences are the additional particle/hole indices on the 
correlation functions $f^{ab}_\mathbf{k}$, 
$g^{abc}_{\mathbf{q}\mathbf{k}}$, and 
$h^{abcd}_{\mathbf{k}\mathbf{p}\mathbf{q}}$, 
as well as the source terms 
$S^{ab}_\mathbf{k}$, 
$S^{abc}_{\mathbf{q}\mathbf{k}}$, and 
$S^{abcd}_{\mathbf{k}\mathbf{p}\mathbf{q}}$. 
Apart from these additional indices, the derivation of the 
Boltzmann equation is completely analogous to the previous case, 
where we finally arrive at 
{\color{black} (see Eq.~\ref{boltzapp})}
\bea
\label{Boltzmann-strong}
&&
\partial_t f_{\mathbf{k}}^d 
=
-2\pi\int_\mathbf{p}\int_\mathbf{q} 
\sum_{a,b,c}
M^{abcd}_\mathbf{p+q,p,k-q,k} 
\times
\nn
&&
\delta
\left(E_\mathbf{p+q}^a-E_\mathbf{p}^b+E_\mathbf{k-q}^c-E_\mathbf{k}^d\right)
\times
\nn
&&
\left[
  f_\mathbf{k}^{d}f_\mathbf{p}^{b}
  \left(1-f_\mathbf{k-q}^{c}\right)\left(1-f_\mathbf{p+q}^{a}\right)
  -
  (f^d_\mathbf{k}f^b_\mathbf{p}\leftrightarrow f^c_\mathbf{k-q}f^a_\mathbf{p+q})
\right].
\nn
\ea
The matrix elements $M^{abcd}_\mathbf{p+q,p,k-q,k}$ contain different 
processes, such as collision of two quasi-particles 
$M^{++++}_\mathbf{p+q,p,k-q,k}$ or two quasi-holes 
$M^{----}_\mathbf{p+q,p,k-q,k}$, 
but also pair-creation processes, e.g., with one incoming quasi-particle 
and two outgoing particles plus one quasi-hole, 
as long as they are allowed by energy conservation -- which is enforced 
by the second line of~\eqref{Boltzmann-strong}. 

{\color{black}
  \subsection{Other geometries}

  The mean field background sketched in Fig.~\ref{figure-schach} is based on 
  a square or hypercubic lattice at half filling. 
  It might be illuminating to discuss other geometries. 
  Since a graphene-type honeycomb lattice is also bipartite, one would get 
  an analogous Mott-type insulator state at half filling 
  (although no longer with a checkerboard structure, of course) 
  by placing the fermions in one sub-lattice and keeping the other one empty.  
  In this case, the dispersion relations $J_\mathbf{k}$ would change, 
  but apart from that the energies $E^\pm_\mathbf{k}$, are still given 
  by the same functional form~\eqref{energies}.

  A triangular lattice, on the other hand, is not bipartite, and thus 
  does not feature such a Mott-type insulator state at half filling.
  However, at a filling factor of one third, one can obtain an analogous 
  state by occupying one sub-lattice (also triangular) while keeping the 
  other two sub-lattices empty (because the triangular lattice is tripartite). 
  In this case, each occupied lattice site would be surrounded by empty sites. 

  Another interesting point is a departure from half-filling.  
  As long as this deviation is small enough such that it does not destroy
  the global checkerboard structure, one could take it into account by 
  an imbalance of the initial conditions for the distribution functions 
  $f^{+}_\mathbf{k}$ and $f^{-}_\mathbf{k}$. 
  Increasing the $f^{+}_\mathbf{k}$ or decreasing $f^{-}_\mathbf{k}$ (i.e., increasing the quasi-hole occupation $1- f^{-}_\mathbf{k}$) corresponds 
  to a filling factor a bit above or below one half, respectively. 

  Even more possibilities may arise if we do not restrict ourselves to a 
  simple nearest-neighbor interaction.
  Depending on the interaction matrix $V_{\mu\nu}$, one obtains a 
  plethora of geometric phases.
  In this context, it is also interesting to note an analogy to 
  ultra-cold atoms in optical lattices (see, e.g., \cite{Polak,Buechler}).  
  In this set-up, it is possible to realize a scenario where the 
  nearest-neighbor interaction of the spinless fermions considered 
  here is replaced by an effective interaction via additional bosonic 
  atoms. 
  This effectively non-local interaction may support new phases such 
  as a super-solid phase, which displays some similarities to the 
  checkerboard structure considered here (although it is not a precise 
  one-to-one correspondence).}

\section{Strong interaction limit}

Let us now consider the strongly interacting limit $V_\mathbf{k}\gg J_\mathbf{k}$ in order to simplify the complicated expressions of the various matrix elements $M^{abcd}_\mathbf{p+q,p,k-q,k}$.
In this regime, the effective band width of the two bands described by Eq.~\eqref{energies} is small compared to the gap between the two because the former scales with $J^2/V$ as compared to the latter scaling with $V$. Collisions between quasi-particles, quasi-holes, or quasi-particle and -hole can, therefore, not provide the energy needed to overcome the gap and create additional particle-hole pairs. 
The dominant process in this limit is the particle-hole scattering which is determined by the matrix elements 
\begin{align}\label{parthole1}
  M_\mathbf{p+q,p,k-q,k}^{0011}&=M_\mathbf{p+q,p,k-q,k}^{1100}\approx \nonumber\\
  V_\mathbf{q} &\left[V_\mathbf{q}+\frac{V_\mathbf{k-q-p}}{V^2(n^\mathcal{A}-n^\mathcal{B})^2}\left(J_\mathbf{p}J_\mathbf{k}+J_\mathbf{p+q}J_\mathbf{k-q}\right)\right]\,.
\end{align}
The seconds term in (\ref{parthole1}) is suppressed by a factor of $J^2/V^2$
since this particle-hole exchange term requires two hopping events.
The particle-particle and hole-hole scattering demands at least four hopping events and is 
given by 
\begin{align}
  M_\mathbf{p+q,p,k-q,k}^{0000}&=
                                 M_\mathbf{p+q,p,k-q,k}^{1111}\approx \frac{V_\mathbf{q}(J_{\mathbf{p+q}}J_{\mathbf{p}}+J_{\mathbf{k-q}}J_{\mathbf{k}})}{V^4(n^\mathcal{A}-n^\mathcal{B})^4}\nonumber\\
  \times &\bigl[V_\mathbf{q}(J_{\mathbf{p+q}}J_{\mathbf{p}}+J_{\mathbf{k-q}}J_{\mathbf{k}}) - \nonumber\\
                               & -V_\mathbf{k-p-q}(J_{\mathbf{p+q}}J_{\mathbf{k}}+J_{\mathbf{k-q}}J_{\mathbf{p}})\bigr]\,.
\end{align}
In addition, there is particle-hole scattering which involves to leading 
order the exchange-term $\sim V_\mathbf{q} V_\mathbf{k-p-q}$,  
\begin{align}\label{parthole2}
  M_\mathbf{p+q,p,k-q,k}^{0110}&=
                                 M_\mathbf{p+q,p,k-q,k}^{1001}\approx \frac{V_\mathbf{q}(J_\mathbf{p+q}J_\mathbf{k-q}+J_\mathbf{p}J_\mathbf{k})}{V^2(n^\mathcal{A}-n^\mathcal{B})^2}\nonumber\\
  \times &\left[\frac{V_\mathbf{q}(J_\mathbf{p+q}J_\mathbf{k-q} + J_\mathbf{p}J_\mathbf{k})}{V^2(n^\mathcal{A}-n^\mathcal{B})^2}+V_\mathbf{k-p-q}\right]\,.
\end{align}
Note that here the contribution $\sim V_\mathbf{q}^2$ is of higher order 
in contrast to the particle-hole scattering channel given by
\eqref{parthole1}.

Keeping the lowest terms only implies a considerable simplification 
of the quantum Boltzmann equation~\eqref{Boltzmann-strong}, which now reads 
\bea
\label{Boltzmann-strong-particle-hole}
&&
\partial_t f_{\mathbf{k}}^+
=
-2\pi\int_\mathbf{p}\int_\mathbf{q} 
V_\mathbf{q}^2\;
\delta\left(E_\mathbf{p+q}^--E_\mathbf{p}^-+E_\mathbf{k-q}^+-E_\mathbf{k}^+\right)
\times
\nn
&&
\left[
  f_\mathbf{k}^{+}f_\mathbf{p}^{-}
  \left(1-f_\mathbf{k-q}^{+}\right)\left(1-f_\mathbf{p+q}^{-}\right)
  -
  (f^+_\mathbf{k}f^-_\mathbf{p}\leftrightarrow f^+_\mathbf{k-q}f^-_\mathbf{p+q})
\right].
\nn
\ea
In this limit of strong interactions, the scattering cross section only depends on the momentum transfer $\mathbf{q}$.

Quasi-particles and quasi-holes have to be considered two distinct classes rather than a pair of particle and antiparticle as in the weakly interacting case.
This becomes clear from the absence of the second term $V_\mathbf{q}V_\mathbf{k-p-q}$ of Eq.~\eqref{smallV} from Eq.~\eqref{Boltzmann-strong-particle-hole}. 
This term is usually interpreted as interference term between processes with exchanged collision partners. 
In the present case, where quasi-particle and quasi-hole are independent, it does appear but is strongly suppressed by the 
denominator 
$V^2  (n^\mathcal{A}-n^\mathcal{B})^2$ in 
the first factor on the right hand side of Eq.~\eqref{parthole2}.

\begin{figure}
 \includegraphics[width=\linewidth]{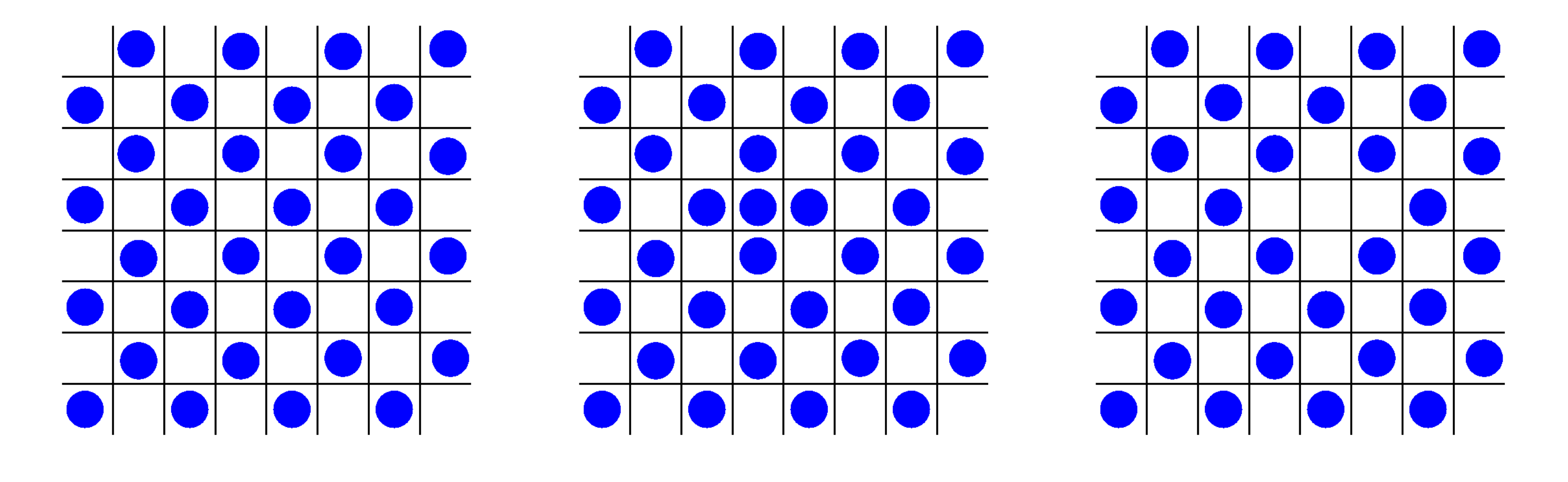}
  \caption{Sketch of a square lattice with a checker-board 
    pattern as an example for a charge-density wave state (left)
    with a quasi-particle (middle) and quasi-hole (right) excitation. 
    By definition of the model, the original Fermions (blue dots) 
    can only move to the nearest neighboring lattice sites,
    i.e., one step in horizontal or vertical direction, 
    but not along the diagonal. 
    Due to the strong repulsion $V$, a quasi-particle (middle) 
    and quasi-hole (right) can only move to next-to-nearest 
    neighboring lattice sites, which involves second-order tunneling 
    processes such as co-tunneling of two Fermions (middle) or 
    sequential tunneling of one Fermion (right).
    We also see that neither two quasi-particles (middle) nor two 
    quasi-holes (right) can occupy nearest neighboring lattice sites. 
  }
  \label{figure-schach}
\end{figure}

Quite intuitively, the suppression of particle-particle (or hole-hole)
collisions can be understood by the observation that two particles cannot 
come close enough to interact directly (same for two holes): they can only 
interact via higher-order virtual hopping processes, see Fig.~\ref{figure-schach}.
In contrast, a quasi-particle and a quasi-hole can occupy neighboring lattice sites 
and thus they can interact directly via $V_{\mu\nu}$. 

As expected, the Boltzmann equations~\eqref{Boltzmann-strong} and 
\eqref{Boltzmann-strong-particle-hole} respect the 
standard conservations laws (e.g., energy, momentum and probability) and 
satisfy the usual consistency conditions (e.g., the crossing relation). 
Note that the quasi-particle $f^+_\mathbf{k}$ and quasi-hole 
$f^-_\mathbf{k}$ excitations obey fermionic statistics, 
consistent with the structure in Eq.s~\eqref{Boltzmann-strong} 
and~\eqref{Boltzmann-strong-particle-hole}, 
i.e., the presence of terms of the type 
$(1-f_\mathbf{k-q}^{+})$ etc. 
As another analogy to the weakly interacting limit~\eqref{smallV},
the quasi-hole distribution function $f^-_\mathbf{k}$ approaches 
unity in the strongly interacting ground state, i.e., the hole 
excitations are properly described by $1-f^-_\mathbf{k}$, as 
in~\eqref{smallV}. 

It is also possible to construct a quantity 
\bea
H=-\sum_a \int_\mathbf{k} \bigl( f_\mathbf{k}^a \ln f_\mathbf{k}^a + (1-f_\mathbf{k}^a) \ln (1-f_\mathbf{k}^a) \bigr)
\ea
that is non-decreasing under collisions and thus to derive an $H$-theorem \cite{Tolman38,B75}. 
As a consequence, the populations of both particles and holes will finally reach stationary distributions, i.e., the system reaches thermalization. 

\section{Time-scale analysis}

Since quasi-particles and quasi-holes are considered to be independent, 
the relaxation described by the quantum Boltzmann equation take 
place on different time scales depending on the preparation of initial conditions.
To explore this possibility, we perform numerical studies on the basis of a specific model: 
The spinless fermions move on a two-dimensional square lattice, 
Coulomb interactions $V$ are limited to nearest-neighbor sites, and $V \gg J$. 
For our calculations we use a ratio of 
{\color{black} $J/V = 10^{-3}$} and set $V$ to one. 
The initial condition is taken as a small perturbation $\delta f$ of the charge-density wave 
\textcolor{black}{that could be realized e.g. by photo-doping~\cite{GBEW19}}.
For this particular choice, the energies entering into the model are given by 
\bea
\label{bands}
E^\pm_\mathbf{k}
=
\frac{V}{2}
\left(
  1\pm\sqrt{1 + \frac{J^{\,2}}{V^2}\,\big(\cos(k_x) + \cos(k_y)\big)^2}
\right)\!.
\ea

Since differences of these energies scale with $J^2/V$ (i.e., the effective band width) 
while the scattering cross section in the Boltzmann equation~\eqref{Boltzmann-strong-particle-hole} 
scales with $V^2$, the typical order of magnitude of the relaxation rate scales with $V^3/J^2$.
A look at the second line of Eq.~\eqref{Boltzmann-strong-particle-hole}, however, 
shows that the rates are also affected by the distribution functions.
To probe this dependence, we vary the initial values for the $f^{\pm}_\mathbf{k}$ 
over a few orders of magnitude and see from our calculations that the relaxation rates 
scale linearly with the initial perturbation $\delta f$.
This behavior can be understood if we look again at Eq.~\eqref{Boltzmann-strong-particle-hole}: 
For a given small perturbation $\delta f \ll 1$ from the charge-density wave ground state 
of the system, we have  $f_\mathbf{k}^+ = \delta f$ and $f_\mathbf{k}^- = 1 -\delta f \approx 1$ 
for the perturbed states.
Inserting into the rate equation \eqref{Boltzmann-strong-particle-hole}, 
we find that the four distribution functions in the rate can be approximated by a 
total factor of $\delta f \, f^+_\mathbf{k}$, which proves the linearity in $\delta f$. 

\begin{figure}[hbt]
  \centering
  \includegraphics[width=\linewidth]{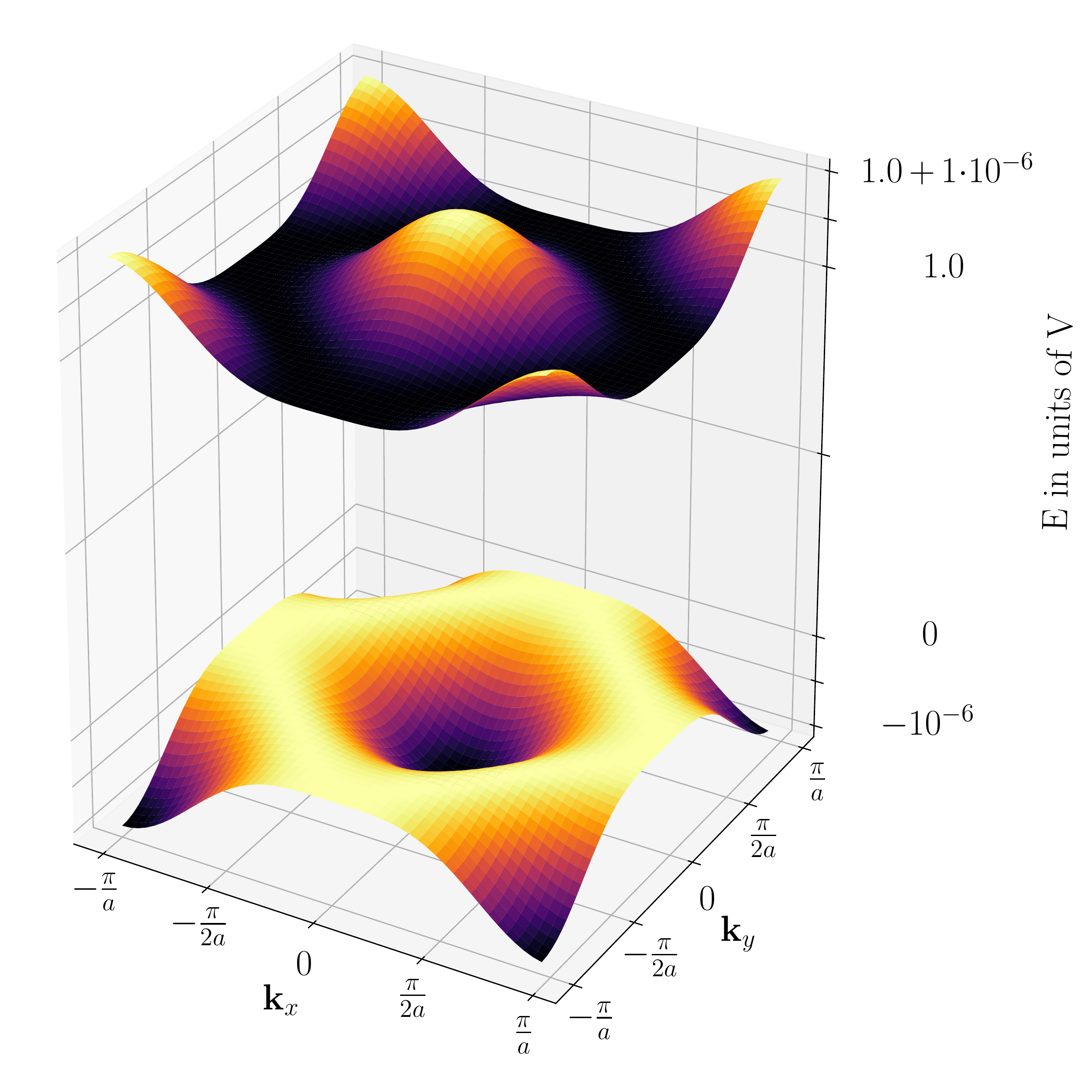}
  \caption{Band structure for a two-dimensional square lattice for $J/V = 10^{-3}$.
    We can see the Brillouin zone and parts of the adjacent ones. 
    The gap is shrunk by six orders of magnitude to show the $\mathbf{k}$-dependence 
    of both bands in one plot.}
  \label{fig:bandstruct}
\end{figure}

Fig.~\ref{fig:bandstruct}\ shows a graphical representation of the band structure.
The upper (quasi-particle) and the lower (quasi-hole) band are mirror-symmetric with 
respect to the center plane of the gap.
The quasi-particle band has a maximum at the center of the Brillouin zone and minima 
at the zone boundaries whereas the lower band is maximal at these k-points and minimal 
at the zone center. 
For simplicity's sake, we will refer to the set of k-points whose eigen-energies 
are closest to the gap as the \emph{diamond}.

Note that the Coulomb matrix element $V_\mathbf{q}$ has a strong $\mathbf{q}$-dependence 
that is equivalent to the $\mathbf{k}$-dependent term under the square root of $E_{\mathbf{k}}$ 
in the case of nearest-neighbor interactions. 
Due to this structure of $V_{\mathbf{q}}$, we have high scattering rates for transitions 
along the diamond (with both initial and final states inside the diamond) as well as for 
transitions between the centers of adjacent Brillouin zones. 
Contrarily, the scattering rates between the zone center and the 
diamond are much reduced and even tend to zero towards the diamond corners.

\begin{figure}[tbph]
  \centering
  \includegraphics[width=.8\linewidth]{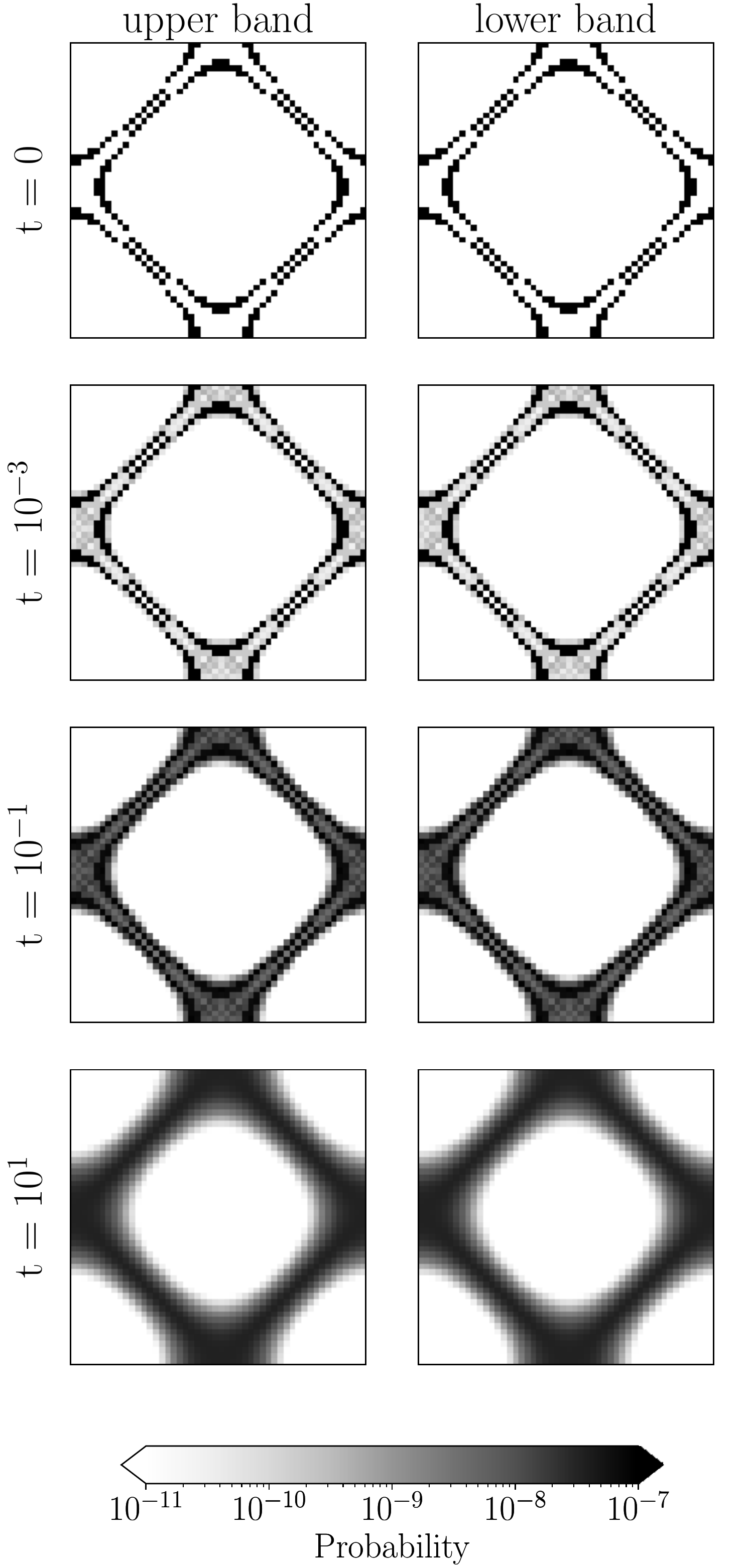}
  \caption{Time series (in units of {\color{black} $J^2/V^3$}) of the evolution of a 
    {\color{black} low-energy} excitation with symmetric initial conditions. 
    For both bands at $t = 0$, a set of states that is close to but not at the gap is perturbed.}
  \label{fig:LowE_time-series}
\end{figure}

\begin{figure}[tbph]
  \centering
  \includegraphics[width=\linewidth]{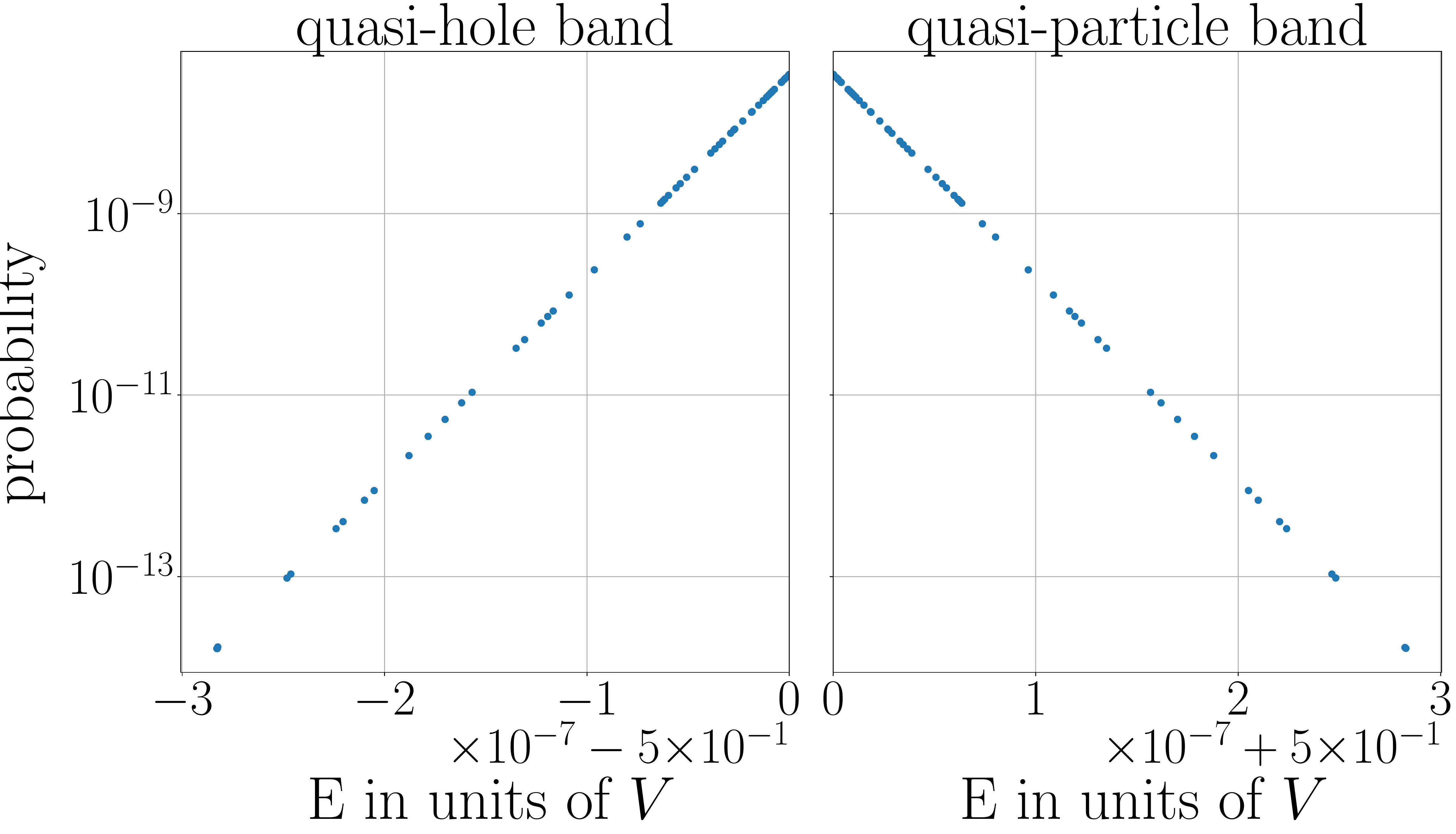}
  \caption{{\color{black} Probability} distribution at {\color{black} $t = 10^{2}J^2/V^3$} 
    for the case of the symmetric initial conditions. 
    Note that the occupation {\color{black} probabilities} for the quasi-holes are 
    {\color{black} plotted} as $1 - f^-$. }
  \label{fig:LowE_E-dist}
\end{figure}

As a first test case, we use initial conditions that are mirror-symmetric with 
respect to the gap and represent a low energy input: 
Quasi-particles and -holes initially occupy the same few $\mathbf{k}$-states 
close to but slightly away from the gap with a low probability of $\delta f = 10^{-7}$ per state.
Given these initial conditions, we then integrate the quantum Boltzmann 
equation on a numerical grid of $50\times50$ k-points in the Brillouin 
zone using an adaptive time step method.

A selected part from the resulting time-series is depicted in Fig.~\ref{fig:LowE_time-series} 
and shows how the distributions evolve. We observe that the scattering among the quasi-particles 
and -holes leads to a spread of their initial distribution over the whole diamond. 
At the end, the states with the lowest (highest for holes) energies have the highest 
occupation probabilities and we see as expected for a thermalized distribution 
that the probabilities decrease towards higher (lower) energies (cf. Fig.~\ref{fig:LowE_E-dist}).

\begin{figure}[tbph]
  \centering
  \includegraphics[width=.8\linewidth]{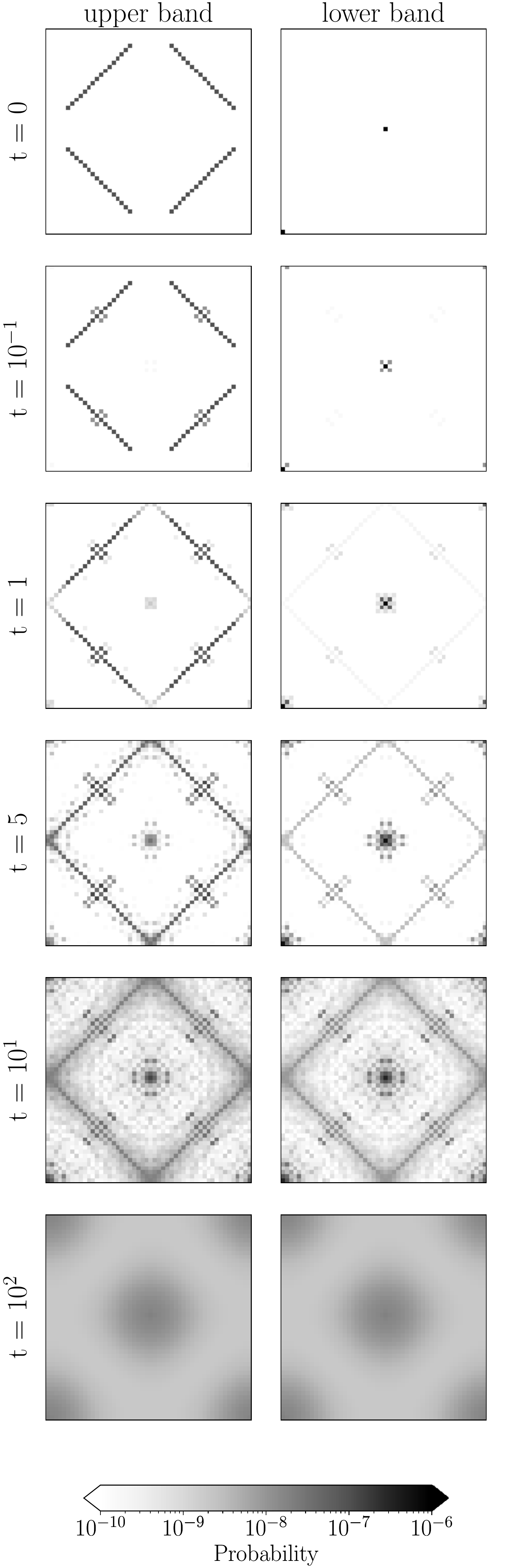}
  \caption{Times series (in units of {\color{black} $J^2/V^3$}) 
    of the evolution of a high energy excitation. 
    The upper (quasi-particle) band is initially perturbed in the diamond shaped 
    {\color{black} minimum}, 
    the lower (quasi-hole) band in the zone center.}
  \label{fig:HighE_time-series}
\end{figure}

\begin{figure}[tbph]
  \centering
  \vspace{12pt}
  \includegraphics[width=\linewidth]{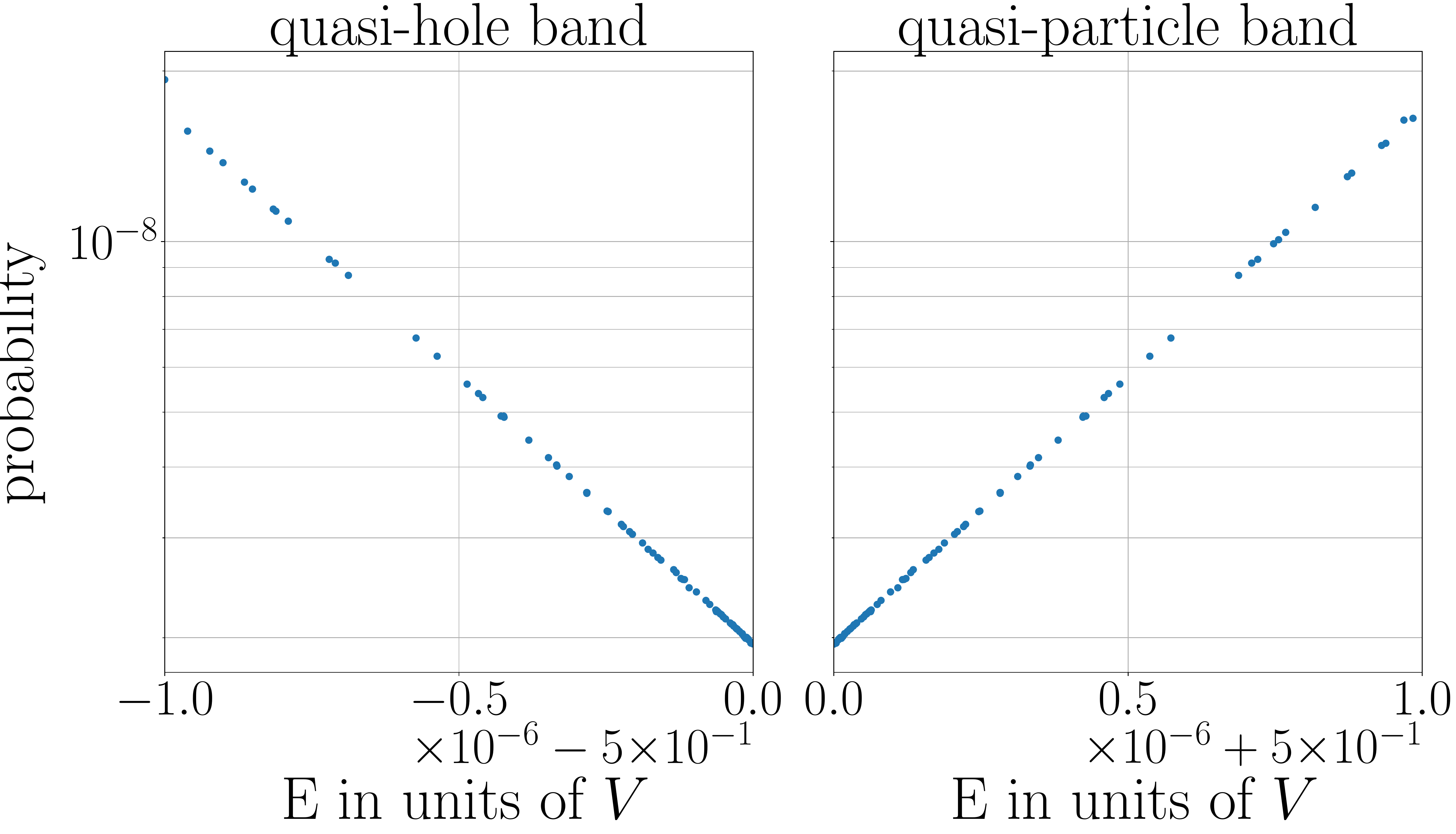}
  \caption{{\color{black} Probability} distribution at {\color{black} $t = 10^{2}J^2/V^3$} 
    for the case of the asymmetric initial conditions. 
    Note that the occupation {\color{black} probabilities} for the {\color{black} quasi-holes are plotted} 
    as $1 - f^-$.}
  \label{fig:HighE_E-dist}
\end{figure}

To explore the consequences of unequal initial populations in the respective bands, 
we choose initial conditions where quasi-particles are again located close to the gap 
while now the holes are located close to the {\color{black} Brillouin} zone center. 
Analyzing the time-series in Fig.~\ref{fig:HighE_time-series}, we notice that the 
relaxation of the initial population proceeds on different time scales.

In the early stages at {\color{black} $t = 10^{-1}J^2/V^3$}, scattering among the holes at 
the center with the particles in the diamond results in a localized broadening around 
the center and the flanks of the diamond. 
At the time {\color{black} $t = J^2/V^3$}, the unoccupied corner states of the diamond 
in the quasi-particle band start to fill. 
Simultaneously, the occupation probabilities of the quasi-hole states close 
to the gap start to increase such that the diamond becomes visible whereas 
for the quasi-particles in turn the center becomes populated.
At later times, the distributions of quasi-particles and quasi-holes within 
the Brillouin zone come to resemble each other more and more closely until 
they eventually become equal at around {\color{black} $t = 10^{2}J^2/V^3$:}  
Now both have their population maximum at the zone center, but the diamond 
is also still populated.

For the interpretation of these results, we have to keep in mind that in the limit $V \gg J$, 
quasi-particles and -holes scatter with each other, but not among themselves. 
A quasi-particle and a quasi-hole are able to exchange both energy and momentum 
in the scattering. 
However, the momentum transfer is governed by the Coulomb matrix element $V_{\mathbf{q}}$, 
assigning different probability to different momentum transfers.

For this reason, we observe in the early stage transitions for which the energy 
is almost conserved separately for quasi-electrons and quasi-holes 
(scattering events from the zone center to an adjacent one or within the diamond by one lattice vector) 
since the momentum dependence of $V_{\mathbf{q}}$ favors these transitions.
At the same time, scattering from the center to the diamond flank with a momentum  
transfer of half a lattice vector contributes as the energy exchange is a good match 
and the Coulomb matrix element is still sizable. 

In the later stages, scattering with arbitrary energy and momentum transfer start to play its role. 
Due to these processes, quasi-particle and -hole occupation probabilities become more and more similar.
The equilibrium configuration for the quasi-particles (quasi-holes) shows a higher (lower) population 
in the high-energy region compared to the diamond region (cf. last panel of Fig.~\ref{fig:HighE_time-series}). 
It corresponds to an inverted Fermi-Dirac distribution characteristic of a system at negative temperature 
as can be seen in the log plot of the distributions in Fig.~\ref{fig:HighE_E-dist}. 
The same applies analogously for the quasi-hole distribution.
Note that such a negative temperature is facilitated by the upper limit for the energy 
(bounded spectrum) and the large initial value for the total energy of the system.

{\color{black}
  \section{Outlook: Back-reaction}
  In the strongly interacting limit $V\gg J$ considered here, the role of the 
  microscopic parameters $J$ and $V$ reduces to a simple overall scaling $V^3/J^2$
  of the relaxation rate (see above), the rest is determined by purely geometrical 
  dimensionless quantities.  Going away from this limit, however, the situation becomes more complex. 

  One the one hand, the eigen-energies $E_\mathbf{k}^\pm$ and the matrix 
  elements $M^{abcd}_\mathbf{p+q,p,k-q,k}$ entering the Boltzmann 
  equation~\eqref{Boltzmann-strong} depend non-trivially on the dimensionless 
  ratio $J/V$. 
  On the other hand, by including back-reaction via 
  $i\partial_t\langle\hat n^{\mathcal A}\rangle
  =
  -i\partial_t\langle\hat n^{\mathcal B}\rangle
  =
  \int_\mathbf{k} J_\mathbf{k}
  \left(
    f^\mathcal{BA}_\mathbf{k}-f^\mathcal{AB}_\mathbf{k}
  \right)$,
  the time-dependence of the distribution functions $f^\pm_\mathbf{k}$
  stemming from the Boltzmann equation~\eqref{Boltzmann-strong}
  does also entail a time-dependence of the sub-lattice fillings 
  $\langle\hat n^{\mathcal A}\rangle$ and $\langle\hat n^{\mathcal B}\rangle$,
  which in turn modify the eigen-energies $E_\mathbf{k}^\pm$ and the matrix 
  elements $M^{abcd}_\mathbf{p+q,p,k-q,k}$ entering the Boltzmann 
  equation~\eqref{Boltzmann-strong} in a time-dependent manner. 

  In principle, these intricate inter-dependencies can be taken in account 
  self-consistently.  
  Fortunately, in the limit of small $J/V\ll1$ and small populations 
  $f^{+}_\mathbf{k}\ll1$ and $f^{-}_\mathbf{k}\approx1$ considered here,
  all these modifications are tiny and can be neglected.
}

\section{Conclusions}
For strongly interacting spinless fermions on a general regular bipartite lattice in higher dimensions, 
we employ the hierarchy of correlations  in order to derive the quasi-particle and quasi-hole 
excitations, their spectrum  
as well as their mutual interactions, which allows us 
to obtain a quantum Boltzmann equation~\cite{arxiv}.  
In the strong-coupling limit, the ground state (at half filling) 
is given by the charge-density wave state, quite analogous to the Mott insulator phase in the Fermi-Hubbard model. 
In this limit, we find that collisions between quasi-particles 
and quasi-holes dominate over particle-particle and hole-hole scattering events. 

As a result, the relaxation and thermalization dynamics strongly depends 
both on the absolute magnitude and on the initial distribution of the 
excitations (quasi-particles or quasi-holes) in the Brillouin zone.
For small initial quasiparticle populations, their lifetime is inversely 
proportional to the initial occupation probability, i.e. the strength of the excitation.
Due to the varying efficiency of momentum transfer, relaxation proceeds 
in two stages if the distributions of quasiparticles
and holes in the Brillouin zone are initially very different. 
Only in the second stage, the distributions of quasiparticles
and holes begin to resemble each other. 
The $H$-theorem for the quantum Boltzmann equation ensures 
that a unique equilibrium state is finally reached.

In summary, we demonstrated that thermalization even in a 
strongly correlated system can still be described in high-dimensional 
systems within the well-known framework of a quantum Boltzmann equation, 
but the solutions of this equation fall into different classes 
depending on the initial conditions chosen.
As possible directions for future work, one could study initial 
conditions departing from half-filling (corresponding to doped Mott insulators) 
or with parameters close to the metal-insulator transition.

\section*{Acknowledgements}
We acknowledge funding by DFG (German Research Foundation), grant 278162697 (SFB 1242) and grant 398912239.

\appendix
\section{Correlators and definitions}

Before we present the details of our calculation, we give here the explicit form of the correlation functions 
and their Fourier representations.
For spinless fermions, the Heisenberg equations for the annihilation and creation operators are
\begin{align}\label{heisenberg}
  i\partial_t \hat{c}_\alpha&=-\frac{1}{Z}\sum_{\mu}J_{\mu\alpha}\hat{c}_\mu+\frac{1}{2Z}\sum_\mu V_{\mu\alpha}(\hat{c}_\alpha \hat n_\mu+ \hat n_\mu\hat{c}_\alpha) \\
  i\partial_t \hat{c}^\dagger_\alpha&=\frac{1}{Z}\sum_{\mu}J_{\mu\alpha}\hat{c}_\mu^\dagger-\frac{1}{2Z}\sum_\mu V_{\mu\alpha}(\hat{c}^\dagger_\alpha  \hat n_\mu+
                                      \hat n_\mu\hat{c}^\dagger_\alpha  )\,. 
\end{align}
Using \eqref{heisenberg}, we can deduce the equation of motion for arbitrary $n$-point expectation values.
Since the hierarchy is based on the correlations among lattice sites, we need in addition 
the relation between $n$-point correlators and $n$-point expectation values.
Up to first order in $1/Z$, we have for $\mu\neq\nu$ the two-point correlations
\begin{align}
  \langle \hat{c}^\dagger_\mu \hat{c}_\nu\rangle^\mathrm{corr}=\int_\mathbf{k} 
  f^\mathrm{corr}_\mathbf{k}e^{i\mathbf{k}\cdot(\mathbf{x}_{\mu}-\mathbf{x}_{\nu})}
\end{align}
and the particle-number correlations $\langle \hat{n}_\mu \hat{n}_\nu\rangle^\mathrm{corr}=
\langle \hat{n}_\mu \hat{n}_\nu\rangle-\langle \hat{n}_\mu \rangle\langle\hat{n}_\nu\rangle$
which will be omitted in the following since they do not contribute to the Boltzmann collision terms in leading order.
The relevant three-point correlators in second order of the hierarchical expansion are given for $\alpha\neq\mu\neq\nu$ by
\begin{align}
  \langle \hat{n}_\alpha\hat{c}^\dagger_\mu \hat{c}_\nu\rangle^\mathrm{corr}&= 
                                                                              \langle \hat{n}_\alpha\hat{c}^\dagger_\mu \hat{c}_\nu\rangle-\langle \hat{n}_\alpha\rangle 
                                                                              \langle\hat{c}^\dagger_\mu \hat{c}_\nu\rangle^\mathrm{corr}
\end{align}
and has the Fourier decomposition
\begin{align}
  \langle \hat n_\alpha\hat c_{\mu}^\dagger \hat c_{\nu}\rangle^\mathrm{corr}&=
                                                                               \int_{\mathbf{p}_1,\mathbf{p}_2} 
                                                                               g_{\mathbf{p}_1,\mathbf{p}_2}e^{i\mathbf{p}_1\cdot(\mathbf{x}_{\mu}-\mathbf{x}_{\alpha})+
                                                                               i\mathbf{p}_2\cdot(\mathbf{x}_{\nu}-\mathbf{x}_{\alpha})}\,.
\end{align}
Furthermore, the 4-point correlators for $\alpha\neq\beta\neq\mu\neq\nu$ are defined as
\begin{align}
  \langle\hat c^\dagger_{\alpha}\hat c_{\beta}\hat c^\dagger_{\mu}\hat c_{\nu}\rangle^\mathrm{corr}=&
                                                                                                      \langle\hat c^\dagger_{\alpha}\hat c_{\beta}\hat c^\dagger_{\mu}\hat c_{\nu}\rangle-
                                                                                                      \langle\hat c^\dagger_{\alpha}\hat c_{\beta}\rangle^\mathrm{corr} \langle\hat c^\dagger_{\mu}\hat c_{\nu}\rangle^\mathrm{corr}\nonumber\\
                                                                                                    &+\langle\hat c^\dagger_{\alpha}\hat c_{\nu}\rangle^\mathrm{corr} \langle\hat c^\dagger_{\mu}\hat c_{\beta}\rangle^\mathrm{corr}
\end{align}
and we define their Fourier components via 
\begin{align}
  \langle\hat c^\dagger_{\alpha}\hat c_{\beta}\hat c^\dagger_{\mu}&\hat c_{\nu}\rangle^\mathrm{corr}
                                                                    =\int_{\mathbf{p}_1,\mathbf{p}_2,\mathbf{p}_3} h_{\mathbf{p}_1,\mathbf{p}_2,\mathbf{p}_3}\nonumber\\
                                                                  & \times e^{i\mathbf{p}_1\cdot(\mathbf{x}_{\alpha}-\mathbf{x}_{\nu})+i\mathbf{p}_2\cdot(\mathbf{x}_{\beta}-
                                                                    \mathbf{x}_{\nu})+i\mathbf{p}_3\cdot(\mathbf{x}_{\mu}-\mathbf{x}_{\nu})}\,.
\end{align}

\section {Homogeneous lattice}
It is instructive to derive with the hierarchical 
method the well-known Boltzmann equations for a homogeneous lattice at half filling, see Eq.~\eqref{smallV}.
The homogeneity of the fermion distribution enforces time-independence of the on-site occupation number which translates to the
zeroth-order equation $\partial_t \langle \hat{n}_\mu\rangle=0$.
The two-point correlators remain constant 
in order $1/Z$ but their equations of motion have 
an inhomogeneity of order $1/Z^2$ which is determined by the three-point correlators,
\begin{align}\label{twopointhom}
  i\partial_t \langle \hat c_\mu^\dagger \hat c_\nu\rangle^\mathrm{corr}&=S^{(2)}_{\mu\nu}=-\frac{1}{Z}\sum_{\alpha}(V_{\alpha\mu}-V_{\alpha\nu})
                                                                          \langle \hat n_\alpha \hat c_\mu^\dagger \hat c_\nu\rangle^\mathrm{corr}\,,
\end{align}
or, translated to Fourier space,
\begin{align}\label{fourier2point}
  i\partial_t f_\mathbf{k}^\mathrm{corr}=S_\mathbf{k}^{(2)}=-\int_\mathbf{q}V_{\mathbf{k}+\mathbf{q}}\left(g_{\mathbf{q},\mathbf{k}}-g_{\mathbf{k},\mathbf{q}}\right)\,.
\end{align}
From the hierarchy of correlations follows the evolution equation for the 
three-point correlators which contains the two-point correlator $\langle \hat c_\mu^\dagger \hat c_\nu\rangle^\mathrm{corr}$ and 
the particle number correlator $\langle \hat n_\mu \hat n_\nu\rangle^\mathrm{corr} $.
As mentioned in the previous section, the latter do not contribute to Boltzmann collisions terms in order $1/Z^3$.
We find
\begin{align}\label{threepointrealspacehom}
  i\partial_t \langle &\hat n_\alpha \hat c_\mu^\dagger \hat c_\nu\rangle^\mathrm{corr}=\nonumber\\ 
  =& 
     \frac{1}{Z}\sum_{\gamma}J_{\gamma\mu}
     \langle \hat n_\alpha \hat c_\gamma^\dagger \hat c_\nu\rangle^\mathrm{corr}
     -\frac{1}{Z}\sum_{\gamma}
     J_{\gamma\nu}\langle \hat n_\alpha \hat c_\mu^\dagger \hat c_\gamma\rangle^\mathrm{corr}\nonumber\\
                      &+S^{(3)}_{\alpha\mu\nu,1/Z^2}+S^{(3)}_{\alpha\mu\nu,1/Z^3}
\end{align}
with the source terms 
\begin{align}\label{inhomZ2hom}
  &S^{(3)}_{\alpha\mu\nu,1/Z^2}=&\nonumber\\
  =&
     \frac{1}{Z}\sum_{\gamma}J_{\gamma\alpha}\left[
     \langle \hat c^\dagger_\alpha \hat c_\nu\rangle^\mathrm{corr} \langle \hat c^\dagger_\mu \hat c_\gamma\rangle^\mathrm{corr}
     -\langle \hat c^\dagger_\gamma \hat c_\nu\rangle^\mathrm{corr} \langle \hat c^\dagger_\mu \hat c_\alpha\rangle^\mathrm{corr}\right]
     \nonumber\\
  &-\frac{1}{4}\left(\frac{V_{\mu\alpha}}{Z}-\frac{V_{\nu\alpha}}{Z}\right)
    \langle \hat c_\mu^\dagger \hat c_\nu \rangle^\mathrm{corr}+ \ldots \,.
\end{align}
and
\begin{align}\label{inhomZ3hom}
  &S^{(3)}_{\alpha\mu\nu,1/Z^3}=\nonumber\\
  &=
    \frac{1}{Z}\sum_{\gamma}J_{\gamma\alpha}\left[\langle \hat c^\dagger_\gamma
    \hat c_\alpha \hat c^\dagger_\mu \hat c_\nu\rangle^\mathrm{corr}
    -\langle \hat c^\dagger_\alpha \hat c_\gamma \hat c^\dagger_\mu \hat c_\nu\rangle^\mathrm{corr}
    \right]+ \ldots \,.
\end{align}
In \eqref{inhomZ2hom} we suppressed the particle-number correlations and in \eqref{inhomZ3hom} we suppressed all terms except the four-point correlators.
As will be shown below, the latter are relevant for the Boltzmann dynamics in leading order.
After the Fourier transformation of \eqref{threepointrealspacehom}, \eqref{inhomZ2hom} and \eqref{inhomZ3hom}, we obtain
\begin{align}\label{Fourierhom3point}
  i\partial_t g_{\mathbf{q},\mathbf{k}}&=(J_{\mathbf{q}}-J_{\mathbf{k}})g_{\mathbf{q},\mathbf{k}}+S^{(3)}_{\mathbf{q},\mathbf{k},1/Z^2}+
                                         S^{(3)}_{\mathbf{q},\mathbf{k},1/Z^3}\,. 
\end{align}
with
\begin{align}
  S^{(3)}_{\mathbf{q},\mathbf{k},1/Z^2}=&(J_{\mathbf{q}}-J_{\mathbf{k}})f^\mathrm{corr}_{\mathbf{q}}f^\mathrm{corr}_{\mathbf{k}}
                                          -\frac{1}{4}V_{\mathbf{q}+\mathbf{k}}
                                          (f^\mathrm{corr}_{\mathbf{k}}-f^\mathrm{corr}_{\mathbf{q}})
\end{align}
and
\begin{align}
  S^{(3)}_{\mathbf{q},\mathbf{k},1/Z^3}=&\int_{\mathbf{p}}
                                          \left(J_{\mathbf{p}}-J_{\mathbf{k+q+p}}\right)h_{\mathbf{p},-\mathbf{k}-\mathbf{q}-\mathbf{p},\mathbf{q}}\,.\label{source3point3}
\end{align}
We integrate the evolution equation \eqref{Fourierhom3point} 
within the Markov approximation and obtain
\begin{align}\label{markovthreepoint}
  g_{\mathbf{q},\mathbf{k}}&=\frac{i(S^{(3)}_{\mathbf{q},\mathbf{k},1/Z^2}+
                             S^{(3)}_{\mathbf{q},\mathbf{k},1/Z^3})}{i(J_\mathbf{k}-J_\mathbf{q})-\epsilon}\,.
\end{align}
\begin{widetext}
  Finally, we have to consider the dynamics of the 4-point correlators 
  which is given in real space by
  \begin{align}\label{fourpointode}
    i\partial_t \langle \hat c^\dagger_\alpha \hat c_\beta \hat c^\dagger_\mu \hat c_\nu\rangle^\mathrm{corr}
    =&\frac{1}{Z}\sum_{\gamma} J_{\gamma\alpha}\langle \hat c^\dagger_\gamma \hat c_\beta \hat c^\dagger_\mu \hat c_\nu\rangle^\mathrm{corr}
       -\frac{1}{Z}\sum_{\gamma} J_{\gamma\beta}\langle \hat c^\dagger_\alpha \hat c_\gamma \hat c^\dagger_\mu \hat c_\nu\rangle^\mathrm{corr}
       +\frac{1}{Z}\sum_{\gamma} J_{\gamma\mu}\langle \hat c^\dagger_\alpha \hat c_\beta \hat c^\dagger_\gamma \hat c_\nu\rangle^\mathrm{corr}\nonumber\\
     &-\frac{1}{Z}\sum_{\gamma} J_{\gamma\nu}\langle \hat c^\dagger_\alpha \hat c_\beta \hat c^\dagger_\mu \hat c_\gamma\rangle^\mathrm{corr}
       +S^{(4)}_{\alpha\beta\mu\nu,1/Z^3}+\mathcal{O}(1/Z^4)
  \end{align}
  with the source term
  \begin{align}\label{4pointsource}
    S^{(4)}_{\alpha\beta\mu\nu,1/Z^3}&=\frac{J_{\alpha\beta}}{Z}\Big[\langle \hat n_\beta \hat c^\dagger_\mu \hat c_\nu\rangle^\mathrm{corr}-
                                       \langle \hat n_\alpha \hat c^\dagger_\mu \hat c_\nu\rangle^\mathrm{corr}
                                       +\langle \hat c^\dagger_\mu \hat c_\beta\rangle^\mathrm{corr}\langle \hat c^\dagger_\beta \hat c_\nu\rangle^\mathrm{corr}
                                       -\langle \hat c^\dagger_\mu \hat c_\alpha\rangle^\mathrm{corr}\langle \hat c^\dagger_\alpha \hat c_\nu\rangle^\mathrm{corr}\Big]\nonumber\\
                                     &+\frac{J_{\alpha\nu}}{Z}\Big[\langle \hat n_\alpha \hat c^\dagger_\mu \hat c_\beta\rangle^\mathrm{corr}-
                                       \langle \hat n_\nu \hat c^\dagger_\mu \hat c_\beta\rangle^\mathrm{corr}
                                       +\langle \hat c^\dagger_\mu \hat c_\alpha\rangle^\mathrm{corr}\langle \hat c^\dagger_\alpha \hat c_\beta\rangle^\mathrm{corr}
                                       -\langle \hat c^\dagger_\mu \hat c_\nu\rangle^\mathrm{corr}\langle \hat c^\dagger_\nu \hat c_\beta\rangle^\mathrm{corr}\Big]\nonumber\\
                                     &+\frac{J_{\beta\mu}}{Z}\Big[\langle n_\mu \hat c^\dagger_\alpha \hat c_\nu\rangle^\mathrm{corr}-
                                       \langle \hat n_\beta \hat c^\dagger_\alpha \hat c_\nu\rangle^\mathrm{corr}
                                       +\langle \hat c^\dagger_\alpha \hat c_\mu\rangle^\mathrm{corr}\langle \hat c^\dagger_\mu \hat c_\nu\rangle^\mathrm{corr}
                                       -\langle \hat c^\dagger_\alpha \hat c_\beta\rangle^\mathrm{corr}\langle \hat c^\dagger_\beta \hat c_\nu\rangle^\mathrm{corr}\Big]\nonumber\\
                                     &+\frac{J_{\mu\nu}}{Z}\Big[\langle \hat n_\nu \hat c^\dagger_\alpha \hat c_\beta\rangle^\mathrm{corr}-
                                       \langle \hat n_\mu \hat c^\dagger_\alpha \hat c_\beta\rangle^\mathrm{corr}
                                       +\langle \hat c^\dagger_\alpha \hat c_\nu\rangle^\mathrm{corr}\langle \hat c^\dagger_\nu \hat c_\beta\rangle^\mathrm{corr}-
                                       \langle \hat c^\dagger_\alpha \hat c_\mu\rangle^\mathrm{corr}\langle \hat c^\dagger_\mu \hat c_\beta\rangle^\mathrm{corr}\Big]\nonumber\\
                                     &-\frac{1}{Z}\sum_\gamma (V_{\alpha\gamma}-V_{\beta\gamma})\langle \hat n_\gamma \hat{c}^\dagger_\mu \hat c_\nu\rangle^\mathrm{corr}
                                       \langle \hat c_\alpha^\dagger \hat c_\beta\rangle^\mathrm{corr}
                                       -\frac{1}{Z}\sum_\gamma (V_{\nu\gamma}-V_{\alpha\gamma})\langle \hat n_\gamma \hat{c}^\dagger_\mu \hat c_\beta\rangle^\mathrm{corr}
                                       \langle \hat c_\alpha^\dagger \hat c_\nu\rangle^\mathrm{corr}\nonumber\\
                                     &-\frac{1}{Z}\sum_\gamma (V_{\beta\gamma}-V_{\mu\gamma})\langle \hat n_\gamma \hat{c}^\dagger_\alpha \hat c_\nu\rangle^\mathrm{corr}
                                       \langle \hat c_\mu^\dagger \hat c_\beta\rangle^\mathrm{corr}
                                       -\frac{1}{Z}\sum_\gamma (V_{\mu\gamma}-V_{\nu\gamma})\langle \hat n_\gamma \hat{c}^\dagger_\alpha \hat c_\beta\rangle^\mathrm{corr}
                                       \langle \hat c_\mu^\dagger \hat c_\nu\rangle^\mathrm{corr}\,.
  \end{align}
  The dynamics of the Fourier components is then governed through
  \begin{align}\label{fourpointFourierapp}
    i\partial_t h_{\mathbf{p}_1,\mathbf{p}_2,\mathbf{p}_3}&=(J_{\mathbf{p}_1}-J_{\mathbf{p}_2}+J_{\mathbf{p}_3}-J_{\mathbf{p}_1+\mathbf{p}_2+\mathbf{p}_3})
                                                            h_{\mathbf{p}_1,\mathbf{p}_2,\mathbf{p}_3}
                                                            +S^{(4)}_{\mathbf{p}_1,\mathbf{p}_2,\mathbf{p}_3,1/Z^3}
  \end{align}
  with
  \begin{align}
    S^{(4)}_{\mathbf{p}_1,\mathbf{p}_2,\mathbf{p}_3,1/Z^3}&=
                                                            (J_{\mathbf{p}_1}-J_{\mathbf{p}_2})f^\mathrm{corr}_{\mathbf{p}_3}f^\mathrm{corr}_{\mathbf{p}_1+\mathbf{p}_2+\mathbf{p}_3}+
                                                            [J_{\mathbf{p}_1+\mathbf{p}_2+\mathbf{p}_3}-
                                                            J_{\mathbf{p}_1}-
                                                            V_{\mathbf{p}_2+\mathbf{p}_3}(f^\mathrm{corr}_{\mathbf{p}_1}-f^\mathrm{corr}_{\mathbf{p}_1+\mathbf{p}_2+\mathbf{p}_3})]g_{\mathbf{p}_3,\mathbf{p}_2}\nonumber\\
                                                          &-(\mathbf{p}_1\leftrightarrow \mathbf{p}_3)+(\mathbf{p}_1\leftrightarrow \mathbf{p}_3,\mathbf{p}_2\leftrightarrow
                                                            -\mathbf{p}_1-\mathbf{p}_2-\mathbf{p}_3)-(\mathbf{p}_2\leftrightarrow -\mathbf{p}_1-\mathbf{p}_2-\mathbf{p}_3)\,.
  \end{align}
  After solving \eqref{fourpointFourierapp} within Markov approximation and plugging the result back into
  (\ref{markovthreepoint}), the evolution equation (\ref{fourier2point}) takes the form
  \begin{align}\label{boltzmann1}
    i\partial_t f^\mathrm{corr}_\mathbf{k}=-\int_\mathbf{q}\frac{iV_{\mathbf{k}+\mathbf{q}}}{i(J_\mathbf{k}-J_\mathbf{q})-\epsilon}
    \left[S^{(3)}_{\mathbf{q},\mathbf{k},1/Z^2}+\int_\mathbf{p}
    \frac{(J_\mathbf{p}-J_{\mathbf{k}+\mathbf{q}+\mathbf{p}})iS^{(4)}_{\mathbf{p},-\mathbf{k}-\mathbf{q}-\mathbf{p},\mathbf{q},1/Z^3}}
    {i(J_\mathbf{k}-J_\mathbf{q}+J_{\mathbf{k}+\mathbf{q}+\mathbf{p}}-J_\mathbf{p})-\epsilon}\right]-c.c.
  \end{align}
  After some algebra and using the identity $\pi\delta(x)=\lim_{\epsilon\rightarrow 0}\epsilon/(\epsilon^2+x^2)$, we find in the continuum limit 
  from (\ref{boltzmann1}) the Boltzmann dynamics
  \begin{align}
    \partial_t f_\mathbf{k}&=
                             -2\pi\int_{\mathbf{q},\mathbf{p}}\delta(J_\mathbf{k}+J_\mathbf{p}-J_\mathbf{k-q}-J_\mathbf{p+q})
                             V_\mathbf{q}(V_\mathbf{q}-V_\mathbf{k-p-q})\nonumber\\
                           &\times\bigg[f_\mathbf{k}f_\mathbf{p}(1-f_\mathbf{k-q})(1-f_\mathbf{p+q})
                             -f_\mathbf{k-q}f_\mathbf{p+q}(1-f_\mathbf{k})(1-f_\mathbf{p})\bigg]\,,
  \end{align}
  where we introduced the electron distribution functions $f_\mathbf{k}$ which are 
  the Fourier components of $\langle\hat{c}^\dagger_\mu \hat{c}_\nu \rangle=\langle\hat{c}^\dagger_\mu \hat{c}_\nu \rangle^\mathrm{corr}
  +\delta_{\mu\nu}\langle\hat{n}_\mu\rangle$, i.e.
  $
  f_\mathbf{k}=1/2+f_\mathbf{k}^\mathrm{corr}$. 
  Finally, we want to remark that in the evaluation of (\ref{boltzmann1}) all terms which do not contribute 
  to the collision terms cancel each other.
  In order to see this, it is necessary to include beside the particle-number correlators 
  also the four-point-correlators 
  $\langle \hat{n}_\alpha \hat{n}_\beta \hat{c}_\mu^\dagger \hat{c}_\nu\rangle^\mathrm{corr} $ 
  (which were not considered in the calculation above) and several local terms which ensure 
  that the correlators vanish identically if two or more lattice sites are equal.

\end{widetext}
\section {Charge-density wave}\label{cd}

\subsection {Single-site evolution.}  We consider a bipartite lattice at half filling such that the 
fermion densities add up to unity, $n^\mathcal{A}+n^\mathcal{B}=1$.
For labeling the sub-lattice we use the capital superscripts such as $X\in \{\mathcal{A},\mathcal{B}\}$.
The time-evolution of the on-site occupation number is given by
\begin{align}\label{singlesitereal}
  i\partial_t\langle \hat n_\mu\rangle=\frac{1}{Z}\sum_\alpha J_{\alpha\mu}\left[\langle \hat{c}^\dagger_\alpha \hat{c}_\mu\rangle^\mathrm{corr}
  -\langle \hat{c}^\dagger_\mu \hat{c}_\alpha\rangle^\mathrm{corr}\right] 
\end{align}
which translates after a Fourier transformation to
\begin{align}\label{onsite}
  i\partial_t n^X=\int_\mathbf{q}J_\mathbf{q}\left[f_\mathbf{q}^{\mathrm{corr},\bar X X}-f_\mathbf{q}^{\mathrm{corr},X\bar X}\right]\,.
\end{align}
The superscript $\bar{X}$ denotes the sub-lattice opposite to $X$. 

\subsection {Quasi-particle and hole distribution functions.}\label{appph}
For the two-point correlations, we generalize the evolution equation (\ref{twopointhom}) for the charge density 
background and find
\begin{align}\label{twopointrealspace}
  i\partial_t \langle \hat c_\mu^\dagger \hat c_\nu\rangle^\mathrm{corr}=&
                                                                           \frac{1}{Z}\sum_{\alpha}J_{\alpha\mu}\langle \hat c_\alpha^\dagger \hat c_\nu\rangle^\mathrm{corr}
                                                                           -\frac{1}{Z}\sum_{\alpha}J_{\alpha\nu}\langle \hat c_\mu^\dagger \hat c_\alpha\rangle^\mathrm{corr}\nonumber\\
                                                                         &-\frac{1}{Z}\sum_{\alpha}(V_{\alpha\mu}-V_{\alpha\nu})\langle \hat n_\alpha\rangle\langle \hat c_\mu^\dagger \hat c_\nu\rangle^\mathrm{corr}\nonumber\\
                                                                         &+S_{\mu\nu,1/Z}+S_{\mu\nu,1/Z^2}
\end{align}
where we separated the source terms according to their order $1/Z$,
\begin{align}
  S_{\mu\nu,1/Z}=&
                   \frac{J_{\mu\nu}}{Z}
                   (\langle \hat n_\nu\rangle-\langle \hat n_\mu\rangle)\nonumber\\
                 &-\delta_{\mu\nu}\frac{1}{Z}\sum_\alpha J_{\alpha\mu}
                   \left[\langle \hat{c}^\dagger_\alpha \hat{c}_\mu\rangle^\mathrm{corr}
                   -\langle \hat{c}^\dagger_\mu \hat{c}_\alpha\rangle^\mathrm{corr}\right]\label{sourcetwopointrealspace1}\,,\\
  S_{\mu\nu,1/Z^2}=&-\frac{1}{Z}\sum_{\alpha}(V_{\alpha\mu}-V_{\alpha\nu})
                     \langle \hat n_\alpha \hat c_\mu^\dagger \hat c_\nu\rangle^\mathrm{corr}\nonumber\\
                 &-\frac{1}{Z}
                   V_{\mu\nu}(\langle \hat{n}_\mu\rangle-\langle \hat{n}_\nu\rangle)\langle \hat{c}^\dagger_\mu \hat{c}_\nu\rangle^\mathrm{corr}\label{sourcetwopointrealspace2}\,.
\end{align}
The second term in (\ref{sourcetwopointrealspace1}) was added such that the evolution equation 
(\ref{twopointrealspace}) is also valid for $\mu=\nu$ and the Fourier summation can 
be performed over all lattice sites.
From (\ref{twopointrealspace}) we find for the evolution of the Fourier components
\begin{align}\label{twopoint0}
  i\partial_t( &f^{\mathrm{corr},XY}_\mathbf{k}+\delta^{XY}n^X)=J_\mathbf{k}(f_\mathbf{k}^{\mathrm{corr},\bar X Y}-f_\mathbf{k}^{\mathrm{corr},X \bar Y })\nonumber\\
               &-(V^{\bar X}-V^{\bar Y})
                 f_\mathbf{k}^{\mathrm{corr},X Y}+S_{\mathbf{k},1/Z}^{X Y}+S_{\mathbf{k},1/Z^2}^{X Y}\,,
\end{align}
where equation (\ref{onsite}) was used.
We can rewrite (\ref{twopoint0}) using the variables
$f^{XY}_\mathbf{k}= f^{\mathrm{corr},XY}_\mathbf{k}+\delta^{XY}n^X $ 
which are the Fourier components of the two-site expectation value 
$\langle\hat{c}^\dagger_\mu \hat{c}_\nu\rangle$,
\begin{align}\label{twopointapp}
  i\partial_t f^{XY}_\mathbf{k}=&J_\mathbf{k}(f_\mathbf{k}^{\bar X Y}-f_\mathbf{k}^{X \bar Y })-(V^{\bar X}-V^{\bar Y})
                                  f_\mathbf{k}^{X Y}\nonumber\\
                                &+S_{\mathbf{k},1/Z}^{X Y}+S_{\mathbf{k},1/Z^2}^{X Y}\,.
\end{align}
The relation (\ref{twopointapp}) can be diagonalized via a rotation in the $X-Y$-subspace by means of 
$f_\mathbf{k}^{ab}=\sum_{XY}O^a_X(\mathbf{k})O^b_Y(\mathbf{k}) f_\mathbf{k}^{XY}$
with the momentum-dependent rotation matrix
\begin{align}
  O_X^{a}(\mathbf{k})=
  \begin{pmatrix}
    \cos\alpha_\mathbf{k}&\sin\alpha_\mathbf{k}\\
    -\sin\alpha_\mathbf{k}&\cos\alpha_\mathbf{k}
  \end{pmatrix}\,.
\end{align}
The entries of this matrix are given by
\begin{align}
  \cos\alpha_\mathbf{k}=\frac{J_\mathbf{k}}{|J_\mathbf{k}|}\frac{\sqrt{\omega_\mathbf{k}+(V^\mathcal{A}-V^\mathcal{B})}}{\sqrt{2\omega_\mathbf{k}}}
\end{align}
and
\begin{align}
  \sin\alpha_\mathbf{k}=\frac{\sqrt{\omega_\mathbf{k}-(V^\mathcal{A}-V^\mathcal{B})}}{\sqrt{2\omega_\mathbf{k}}}
\end{align}
with the eigenfrequency $\omega_\mathbf{k}=\sqrt{(V^\mathcal{A}-V^\mathcal{B})^2+4 J_\mathbf{k}^2}$.
For a slowly varying charge-density background we can assume $[\partial_t,O_X^{a}(\mathbf{k})]\approx 0$ such that 
the diagonalization of~(\ref{twopointapp}) leads to
\begin{align}\label{corrdyn}
  i\partial_t f^{ab}_\mathbf{k}&=(-E^a_\mathbf{k}+E^b_\mathbf{k})
                                 f_\mathbf{k}^{ab}+S_{\mathbf{k},1/Z}^{ab}+S_{\mathbf{k},1/Z^2}^{ab}
\end{align}
with the quasi-particle ($a=+$) and hole ($a=-$) energies $E^\pm_\mathbf{k}=\left[V\pm \omega_\mathbf{k}\right]/2$.
For $a=b$ the variables are the distribution functions for quasi-particles and holes, namely
\begin{align}\label{distribution}
  f^{aa}_\mathbf{k}=f^{\mathrm{corr},aa}_\mathbf{k}+\sum_X O_X^a(\mathbf{k}) O_X^a(\mathbf{k}) n^X\equiv f^{a}_\mathbf{k}\,. 
\end{align}
For the slowly varying distribution functions $f^{a}_\mathbf{k} $, the $1/Z$ source term in (\ref{corrdyn}) 
is vanishing.
Thus, their time evolution is governed by terms which are at least of order $1/Z^2$:
\begin{align}\label{dist1}
  i\partial_t f^{a}_\mathbf{k}=S_{\mathbf{k},1/Z^2}^{aa}\,. 
\end{align}
There are two important identities which are useful for the transformation from the sub-lattice space to the particle-hole space.
The first one is the inversion of equation (\ref{distribution})
\begin{align}\label{inverse}
  f_\mathbf{k}^{\mathrm{corr},XY}=\sum_a O_X^a(\mathbf{k})O_X^a(\mathbf{k})f_\mathbf{k}^{a}-\delta^{XY}n^X+\mathcal{O}(1/Z^2)\,.
\end{align}
which can be derived from the fact that the off-diagonal correlations approach their prethermalized 
value to lowest order, i.e.
\begin{align}
  f^{\mathrm{corr},a\bar{a}}_\mathbf{k}=-\sum_X O_a^X(\mathbf{k})O_{\bar a}^X(\mathbf{k}) n^X+\mathcal{O}(1/Z^2)\,. 
\end{align}
The second identity is the eigenvalue equation for rotation matrix
\begin{align}
  J_\mathbf{k}O_{a}^{X}(\mathbf{k})&=(-E_\mathbf{k}^a-V^{\bar X})O_{a}^{\bar X}(\mathbf{k})\,. 
\end{align}

\subsection {Three-point correlators.} 

The Boltzmann collisions are contained in the $1/Z^2$-term in equation (\ref{twopointapp})
which have the form
\begin{align}
  S_{\mathbf{k},1/Z^2}^{XY}=-\int_\mathbf{q}V_{\mathbf{k}+\mathbf{q}}\left(g^{\bar{X}XY}_{\mathbf{q},\mathbf{k}}-g^{\bar{Y}XY}_{\mathbf{k},\mathbf{q}}\right)\,.
\end{align}
Transforming this source term to particle-hole space, we find from (\ref{dist1}) the generalization
of (\ref{fourier2point}) to be
\begin{align}\label{corr2point}
  i\partial_t f^{a}_\mathbf{k}=-\int_\mathbf{q}\sum_{b,X}V_{\mathbf{k}+\mathbf{q}}\left
  (O_X^a(\mathbf{k})O_X^b(\mathbf{q})g^{\bar{X},ba}_{\mathbf{q,k}}-c.c.\right)\,.
\end{align}
Here we rotated the three-point correlations according to $g^{Z,ab}_{\mathbf{q,k}}=\sum_{X,Y}O_X^a(\mathbf{q})O_Y^b(\mathbf{k})g^{ZXY}_{\mathbf{q,k}}$.
Their dynamics is determined by the real space equation 
\begin{widetext}
  \begin{align}\label{threepointrealspace}
    i\partial_t \langle \hat n_\alpha \hat c_\mu^\dagger \hat c_\nu\rangle^\mathrm{corr}&= 
                                                                                          \frac{1}{Z}\sum_{\gamma}J_{\gamma\mu}
                                                                                          \langle \hat n_\alpha \hat c_\gamma^\dagger \hat c_\nu\rangle^\mathrm{corr} 
                                                                                          -\frac{1}{Z}\sum_{\gamma}
                                                                                          J_{\gamma\nu}\langle \hat n_\alpha \hat c_\mu^\dagger \hat c_\gamma\rangle^\mathrm{corr}
                                                                                          -\frac{1}{Z}\sum_{\gamma}(V_{\gamma\mu}-V_{\gamma\nu})\langle \hat n_\gamma\rangle \langle \hat n_\alpha \hat c_\mu^\dagger \hat c_\nu\rangle^\mathrm{corr}\nonumber\\
                                                                                        &+S_{\alpha\mu\nu,1/Z^2}+S_{\alpha\mu\nu,1/Z^3}
  \end{align}
  which is a generalization of (\ref{threepointrealspacehom}).
  Again, the particle-number correlators can be omitted in the source terms since they contribute 
  with terms that are $\mathcal{O}(1/Z^4)$ to the Boltzmann dynamics whereas we shall see that the leading order 
  collision terms are $\mathcal{O}(1/Z^3)$.
  Therefore we remain with
  \begin{align}\label{inhomZ2}
    S_{\alpha\mu\nu,1/Z^2}&=
                            \frac{1}{Z}\sum_{\gamma}J_{\gamma\alpha}\left[
                            \langle \hat c^\dagger_\alpha \hat c_\nu\rangle^\mathrm{corr} \langle \hat c^\dagger_\mu \hat c_\gamma\rangle^\mathrm{corr}
                            -\langle \hat c^\dagger_\gamma \hat c_\nu\rangle^\mathrm{corr} \langle \hat c^\dagger_\mu \hat c_\alpha\rangle^\mathrm{corr}\right]
                            -\left(\frac{V_{\mu\alpha}}{Z}-\frac{V_{\nu\alpha}}{Z}\right)\langle \hat n_\alpha\rangle
                            (1-\langle \hat n_\alpha\rangle)
                            \langle \hat c_\mu^\dagger \hat c_\nu \rangle^\mathrm{corr}
                            \nonumber\\
                          &+\frac{J_{\alpha\mu}}{Z}\left[(\langle \hat n_\mu\rangle-\langle \hat n_\alpha\rangle)\langle \hat c^\dagger_\alpha 
                            \hat c_\nu\rangle^\mathrm{corr}\right]-\frac{J_{\alpha\nu}}{Z}\left[(\langle \hat n_\nu\rangle-\langle \hat n_\alpha\rangle)
                            \langle \hat c^\dagger_\mu \hat c_\alpha\rangle^\mathrm{corr}\right]+...\,.
  \end{align}
  Within the source term $S_{\alpha\mu\nu,1/Z^3}$, only the four-point correlators are of interest,
  \begin{align}\label{inhomZ3}
    S_{\alpha\mu\nu,1/Z^3}&=
                            \frac{1}{Z}\sum_{\gamma}J_{\gamma\alpha}\left[\langle \hat c^\dagger_\gamma
                            \hat c_\alpha \hat c^\dagger_\mu \hat c_\nu\rangle^\mathrm{corr}
                            -\langle \hat c^\dagger_\alpha \hat c_\gamma \hat c^\dagger_\mu \hat c_\nu\rangle^\mathrm{corr}
                            \right]+...\,.
  \end{align}
  After Fourier transformation and rotation in sub-lattice space, we find from (\ref{threepointrealspace}), (\ref{inhomZ2}) and (\ref{inhomZ3})
  the generalization of the evolution equation (\ref{Fourierhom3point}), i.e.
  \begin{align}\label{threepointdiag}
    i\partial_t g_{\mathbf{q},\mathbf{k}}^{Xab}&=(-E^{a}_{\mathbf{q}}+E^b_{\mathbf{k}})g_{\mathbf{q},\mathbf{k}}^{Xab}
                                                 +S_{\mathbf{q},\mathbf{k},1/Z^2}^{Xab}+S_{\mathbf{q},\mathbf{k},1/Z^3}^{Xab}
  \end{align}
  with
  \begin{align}
    S_{\mathbf{q},\mathbf{k},1/Z^2}^{ X ab}&=(E_{\mathbf{q}}^a-E_{\mathbf{k}}^b) O^a_X(\mathbf{q})O^b_X(\mathbf{k})
                                             \left[-(n^X)^2+n^X(f_{\mathbf{q}}^a+f_{\mathbf{k}}^b)-f_{\mathbf{q}}^af_{\mathbf{k}}^b\right]
                                             -V_{\mathbf{q}+\mathbf{k}}O^{a}_{\bar X}(\mathbf{q})O^{b}_{\bar X}(\mathbf{k})
                                             (n^X-1)n^X(f_{\mathbf{q}}^a-f_{\mathbf{k}}^b)
  \end{align}
  and
  \begin{align}
    S_{\mathbf{q},\mathbf{k},1/Z^3}^{ X ab}&=\int_\mathbf{p}\sum_{c,d}(E_{\mathbf{k+q+p}}^d-E_\mathbf{p}^c)
                                             O_X^c(\mathbf{p})O_X^d(\mathbf{k+q+p})h^{cdab}_{\mathbf{p},\mathbf{-k-q-p},\mathbf{q}}
  \end{align}
  where we rotated the four-point correlators according to
  \begin{align}
    h^{abcd}_{\mathbf{p}_1,\mathbf{p}_2,\mathbf{p}_3}=\sum_{XYVW}O_X^a(\mathbf{p}_1)
    O_Y^b(\mathbf{p}_2)O_V^c(\mathbf{p}_3)O_W^d(\mathbf{p}_1+\mathbf{p}_2+\mathbf{p}_3)
    h^{XYVW}_{\mathbf{p}_1,\mathbf{p}_2,\mathbf{p}_3}\,.
  \end{align}
  \subsection {Four-point correlators.} 

  The dynamics of the Fourier components $h^{XYVW}_{\mathbf{p}_1,\mathbf{p}_2,\mathbf{p}_3}$ 
  can be deduced from a generalization of (\ref{fourpointode}), i.e.
  \begin{align}\label{4pointode}
    i\partial_t \langle \hat c^\dagger_\alpha \hat c_\beta \hat c^\dagger_\mu \hat c_\nu\rangle^\mathrm{corr}
    &=\frac{1}{Z}\sum_{\gamma} J_{\gamma\alpha}\langle \hat c^\dagger_\gamma \hat c_\beta \hat c^\dagger_\mu \hat c_\nu\rangle^\mathrm{corr}
      -\frac{1}{Z}\sum_{\gamma} J_{\gamma\beta}\langle \hat c^\dagger_\alpha \hat c_\gamma \hat c^\dagger_\mu \hat c_\nu\rangle^\mathrm{corr}\nonumber\\
    &+\frac{1}{Z}\sum_{\gamma} J_{\gamma\mu}\langle \hat c^\dagger_\alpha \hat c_\beta \hat c^\dagger_\gamma \hat c_\nu\rangle^\mathrm{corr}
      -\frac{1}{Z}\sum_{\gamma} J_{\gamma\nu}\langle \hat c^\dagger_\alpha \hat c_\beta \hat c^\dagger_\mu \hat c_\gamma\rangle^\mathrm{corr}\nonumber\\
    &-\frac{1}{Z}\sum_\gamma(V_{\alpha\gamma}-V_{\beta\gamma}+V_{\mu\gamma}-V_{\nu\gamma})\langle\hat{n}_\gamma\rangle
      \langle \hat{c}^\dagger_\alpha\hat{c}_\beta\hat{c}^\dagger_\mu\hat{c}_\nu\rangle^\mathrm{corr}+S_{\alpha\beta\mu\nu,1/Z^3}+\mathcal{O}(1/Z^4)\,.
  \end{align}
  The inhomogeneity
  \begin{align}\label{source4point}
    S_{\alpha\beta\mu\nu,1/Z^3}&=\frac{J_{\alpha\beta}}{Z}\Big[\langle \hat n_\beta \hat c^\dagger_\mu \hat c_\nu\rangle^\mathrm{corr}-
                                 \langle \hat n_\alpha \hat c^\dagger_\mu \hat c_\nu\rangle^\mathrm{corr}
                                 +\langle \hat c^\dagger_\mu \hat c_\beta\rangle^\mathrm{corr}\langle \hat c^\dagger_\beta \hat c_\nu\rangle^\mathrm{corr}
                                 -\langle \hat c^\dagger_\mu \hat c_\alpha\rangle^\mathrm{corr}\langle \hat c^\dagger_\alpha \hat c_\nu\rangle^\mathrm{corr}\Big]\nonumber\\
                               &+\frac{J_{\alpha\nu}}{Z}\Big[\langle \hat n_\alpha \hat c^\dagger_\mu \hat c_\beta\rangle^\mathrm{corr}-
                                 \langle \hat n_\nu \hat c^\dagger_\mu \hat c_\beta\rangle^\mathrm{corr}
                                 +\langle \hat c^\dagger_\mu \hat c_\alpha\rangle^\mathrm{corr}\langle \hat c^\dagger_\alpha \hat c_\beta\rangle^\mathrm{corr}
                                 -\langle \hat c^\dagger_\mu \hat c_\nu\rangle^\mathrm{corr}\langle \hat c^\dagger_\nu \hat c_\beta\rangle^\mathrm{corr}\Big]\nonumber\\
                               &+\frac{J_{\beta\mu}}{Z}\Big[\langle n_\mu \hat c^\dagger_\alpha \hat c_\nu\rangle^\mathrm{corr}-
                                 \langle \hat n_\beta \hat c^\dagger_\alpha \hat c_\nu\rangle^\mathrm{corr}
                                 +\langle \hat c^\dagger_\alpha \hat c_\mu\rangle^\mathrm{corr}\langle \hat c^\dagger_\mu \hat c_\nu\rangle^\mathrm{corr}
                                 -\langle \hat c^\dagger_\alpha \hat c_\beta\rangle^\mathrm{corr}\langle \hat c^\dagger_\beta \hat c_\nu\rangle^\mathrm{corr}\Big]\nonumber\\
                               &+\frac{J_{\mu\nu}}{Z}\Big[\langle \hat n_\nu \hat c^\dagger_\alpha \hat c_\beta\rangle^\mathrm{corr}-
                                 \langle \hat n_\mu \hat c^\dagger_\alpha \hat c_\beta\rangle^\mathrm{corr}
                                 +\langle \hat c^\dagger_\alpha \hat c_\nu\rangle^\mathrm{corr}\langle \hat c^\dagger_\nu \hat c_\beta\rangle^\mathrm{corr}-
                                 \langle \hat c^\dagger_\alpha \hat c_\mu\rangle^\mathrm{corr}\langle \hat c^\dagger_\mu \hat c_\beta\rangle^\mathrm{corr}\Big]\nonumber\\
                               &-\frac{1}{Z}\sum_\gamma (V_{\alpha\gamma}-V_{\beta\gamma})\langle \hat n_\gamma \hat{c}^\dagger_\mu \hat c_\nu\rangle^\mathrm{corr}
                                 \langle \hat c_\alpha^\dagger \hat c_\beta\rangle^\mathrm{corr}
                                 -\frac{1}{Z}\sum_\gamma (V_{\nu\gamma}-V_{\alpha\gamma})\langle \hat n_\gamma \hat{c}^\dagger_\mu \hat c_\beta\rangle^\mathrm{corr}
                                 \langle \hat c_\alpha^\dagger \hat c_\nu\rangle^\mathrm{corr}\nonumber\\
                               &-\frac{1}{Z}\sum_\gamma (V_{\beta\gamma}-V_{\mu\gamma})\langle \hat n_\gamma \hat{c}^\dagger_\alpha \hat c_\nu\rangle^\mathrm{corr}
                                 \langle \hat c_\mu^\dagger \hat c_\beta\rangle^\mathrm{corr}
                                 -\frac{1}{Z}\sum_\gamma (V_{\mu\gamma}-V_{\nu\gamma})\langle \hat n_\gamma \hat{c}^\dagger_\alpha \hat c_\beta\rangle^\mathrm{corr}
                                 \langle \hat c_\mu^\dagger \hat c_\nu\rangle^\mathrm{corr}\nonumber\\
                               &-\frac{V_{\alpha\beta}}{Z}(\langle \hat n_\beta\rangle-\langle \hat n_\alpha\rangle)
                                 \langle \hat c^\dagger_\alpha \hat c_\nu\rangle^\mathrm{corr}\langle \hat c^\dagger_\mu \hat c_\beta\rangle^\mathrm{corr}
                                 -\frac{V_{\mu\nu}}{Z}(\langle \hat n_\nu\rangle-\langle \hat n_\mu\rangle)
                                 \langle \hat c^\dagger_\alpha \hat c_\nu\rangle^\mathrm{corr}\langle \hat c^\dagger_\mu \hat c_\beta\rangle^\mathrm{corr}\nonumber\\
                               &-\frac{V_{\alpha\nu}}{Z}(\langle \hat n_\alpha\rangle-\langle \hat n_\nu\rangle)
                                 \langle \hat c^\dagger_\alpha \hat c_\beta\rangle^\mathrm{corr}\langle \hat c^\dagger_\mu \hat c_\nu\rangle^\mathrm{corr}
                                 -\frac{V_{\beta\mu}}{Z}(\langle \hat n_\mu\rangle-\langle \hat n_\beta\rangle)
                                 \langle \hat c^\dagger_\alpha \hat c_\beta\rangle^\mathrm{corr}\langle \hat c^\dagger_\mu \hat c_\nu\rangle^\mathrm{corr}
  \end{align}
  contains additional terms compared to (\ref{4pointsource}) due to the presence of the charge density wave.
  A transformation of (\ref{4pointode}) and (\ref{source4point}) to Fourier space and a subsequent rotation 
  in sub-lattice space leads to
  \begin{align}\label{fourpoint}
    i\partial_t h^{abcd}_{\mathbf{p}_1,\mathbf{p}_2,\mathbf{p}_3}=
    (-E_{\mathbf{p}_1}^a+E_{\mathbf{p}_2}^b-E_{\mathbf{p}_3}^c+E_{\mathbf{p}_1+\mathbf{p}_2+\mathbf{p}_3}^d)h^{abcd}_{\mathbf{p}_1,\mathbf{p}_2,\mathbf{p}_3}
    +S^{abcd}_{\mathbf{p}_1,\mathbf{p}_2,\mathbf{p}_3,1/Z^3}+\mathcal{O}(1/Z^4)
  \end{align}
  with
  \begin{align}
    &S^{abcd}_{\mathbf{p}_1,\mathbf{p}_2,\mathbf{p}_3,1/Z^3}\nonumber\\
    =&
       \sum_X O_X^a(\mathbf{p}_1)O_{\bar{X}}^b (\mathbf{p}_2)O_{\bar{X}}^c(\mathbf{p}_3)
       O_X^d(\mathbf{p}_1+\mathbf{p}_2+\mathbf{p}_3)V_{\mathbf{p}_2+\mathbf{p}_3}(n^{\bar{X}}-n^X)\left[f_{\mathbf{p}_3}^c f_{\mathbf{p}_1+\mathbf{p}_2+\mathbf{p}_3}^d-n^X f_{\mathbf{p}_3}^c-
       n^{\bar{X}} f_{\mathbf{p}_1+\mathbf{p}_2+\mathbf{p}_3}^d\right]\nonumber\\
    +&\sum_X O_X^a(\mathbf{p}_1)O_{X}^b (\mathbf{p}_2)O_{X}^c(\mathbf{p}_3)
       O_X^d(\mathbf{p}_1+\mathbf{p}_2+\mathbf{p}_3)\Big[\left(E^b_{\mathbf{p}_2}-E^a_{\mathbf{p}_1}\right)
       f^c_{\mathbf{p}_3}f^d_{\mathbf{p}_1+\mathbf{p}_2+\mathbf{p}_3}-n^X E_{\mathbf{p}_3}^c f^d_{\mathbf{p}_1+\mathbf{p}_2+\mathbf{p}_3}
       +n^X E_{\mathbf{p}_1+\mathbf{p}_2+\mathbf{p}_3}^d f^c_{\mathbf{p}_3}\Big]\nonumber\\
    +&\sum_X g^{Xcb}_{\mathbf{p}_3,\mathbf{p}_2}\Big[\left(E^a_{\mathbf{p}_1}-E^d_{\mathbf{p}_1+\mathbf{p}_2+\mathbf{p}_3}\right)
       O_X^a(\mathbf{p}_1)O_X^d(\mathbf{p}_1+\mathbf{p}_2+\mathbf{p}_3)-V_{\mathbf{p}_2+\mathbf{p}_3}O_X^a(\mathbf{p}_1)O_X^d(\mathbf{p}_1+\mathbf{p}_2+\mathbf{p}_3)\left(f^a_{\mathbf{p}_1}-n^X\right)\nonumber\\
    &+V_{\mathbf{p}_2+\mathbf{p}_3}O_{\bar X}^a(\mathbf{p}_1)O_{\bar X}^d(\mathbf{p}_1+\mathbf{p}_2+\mathbf{p}_3)
      \left(f^d_{\mathbf{p}_1+\mathbf{p}_2+\mathbf{p}_3}-n^{\bar X}\right)\Big]\nonumber\\
    &-(\{a ,\mathbf{p}_1\}\leftrightarrow \{c,\mathbf{p}_3\})+(\{a ,\mathbf{p}_1\}\leftrightarrow \{c,\mathbf{p}_3\},
      \{b,\mathbf{p}_2\}\leftrightarrow\{ d,-\mathbf{p}_1-\mathbf{p}_2-\mathbf{p}_3\})-(\{b,\mathbf{p}_2\}\leftrightarrow\{ d,-\mathbf{p}_1-\mathbf{p}_2-\mathbf{p}_3\})\,.
  \end{align}
  \subsection {Boltzmann dynamics.} As in the previous section, the differential equations for the three-point correlators (\ref{threepointdiag}) and 
  the four-point correlators (\ref{fourpoint}) are solved within Markov approximation.
  When the resulting expressions are inserted into the evolution equation for the 
  particle and hole distribution functions (\ref{corr2point}) we find
  \begin{align}
    i\partial_t f^a_\mathbf{k}=&-\int_{\mathbf{q}}\sum_{X,b}V_{\mathbf{k}+\mathbf{q}}
                                 \frac{iO_X^b(\mathbf{q})O_X^a(\mathbf{k})}{i(E^b_\mathbf{q}-E^a_\mathbf{k})-\epsilon}
                                 \Bigg[S^{\bar{X},ba}_{\mathbf{q},\mathbf{k},1/Z^2}\nonumber\\
                               &+\int_\mathbf{p}\sum_{c,d}(E^d_{\mathbf{k}+\mathbf{q}+\mathbf{p}}-E^c_\mathbf{p})
                                 \frac{i O_X^c(\mathbf{p})O_X^d(\mathbf{k}+\mathbf{q}+\mathbf{p})S^{cdba}_{\mathbf{p},-\mathbf{k}-\mathbf{q}-\mathbf{p},\mathbf{q},1/Z^3}}
                                 {i(E^c_\mathbf{p}-E^d_{\mathbf{k}+\mathbf{q}+\mathbf{p}}+E^b_\mathbf{q}-E^a_\mathbf{k})-\epsilon}\Bigg]-c.c.
  \end{align}
  After some algebra and taking the continuum limit, one can show that the Boltzmann equations take the form
  \begin{align}\label{boltzapp}
    \partial_t f_{\mathbf{k}}^d =& -2\pi\int_{\mathbf{q,p}}\sum_{a,b,c}M^{abcd}_\mathbf{p+q,p,k-q,k} 
                                   \delta(E_\mathbf{p+q}^a-E_\mathbf{p}^b+E_\mathbf{k-q}^c-E_\mathbf{k}^d)\nonumber\\
    \quad&\times\Big[f_\mathbf{k}^{d}f_\mathbf{p}^{b}(1-f_\mathbf{k-q}^{c})(1-f_\mathbf{p+q}^{a})-f_\mathbf{p+q}^{a}f_\mathbf{p-q}^{c}
           (1-f_\mathbf{k}^{d})(1-f_\mathbf{p}^{b})\Big]
  \end{align}
  with the transition matrix elements given by
  \begin{align}
    M_\mathbf{p+q,p,k-q,k}^{abcd}&=\sum_{X,Y}V_\mathbf{q}O^{a}_X(\mathbf{p+q})O^{b}_X(\mathbf{p})
                                   O^{c}_{ \bar X}(\mathbf{k-q})O^{d}_{ \bar X}(\mathbf{k})\nonumber\\
                                 &\times
                                   \left[V_\mathbf{q}O^{a}_{Y}(\mathbf{p+q})O^{b}_{Y}(\mathbf{p})O^{c}_{ \bar Y}(\mathbf{k-q})O^{d}_{ \bar Y}(\mathbf{k})
                                   -V_\mathbf{k-p-q}O^{a}_{ Y}(\mathbf{p+q})O^{b}_{\bar  Y}(\mathbf{p})O^{c}_{ \bar Y}(\mathbf{k-q})O^{d}_{ Y}(\mathbf{k})\right]\,.
  \end{align}

  \subsection {Charge density background.} 
  The collision dynamics has also an impact on the charge density background.
  We know from (\ref{onsite}) that the change of the local charge density is determined 
  by the off-diagonal correlation functions $f^{\mathrm{corr},a\bar{a}}_\mathbf{k}$.
  From the relation (\ref{corrdyn}) we find that their dynamics is determined 
  by the Boltzmann collisions of the particle and hole distribution functions $f_\mathbf{k}^a$.
  After some algebra one arrives at the result
  \begin{align}
    \partial_t n^\mathcal{A}=-\partial_t n^\mathcal{B} =& -2\pi\int_{\mathbf{k,q,p}}\sum_{a,b,c,d}N^{abcd}_\mathbf{p+q,p,k-q,k} 
                                                          \delta(E_\mathbf{p+q}^a-E_\mathbf{p}^b+E_\mathbf{k-q}^c-E_\mathbf{k}^d)\nonumber\\
    \quad&\times\Big[f_\mathbf{k}^{d}f_\mathbf{p}^{b}(1-f_\mathbf{k-q}^{c})(1-f_\mathbf{p+q}^{a})-f_\mathbf{p+q}^{a}f_\mathbf{p-q}^{c}
           (1-f_\mathbf{k}^{d})(1-f_\mathbf{p}^{b})\Big]
           \label{eq:nAB}
  \end{align}
  with 
  \begin{align}
    N_\mathbf{p+q,p,k-q,k}^{abcd}&=\frac{J_\mathbf{k}}{\omega_\mathbf{k}}\sum_{X,Y}V_\mathbf{q}O^{a}_X(\mathbf{p+q})O^{b}_X(\mathbf{p})
                                   O^{c}_{ \bar X}(\mathbf{k-q})O^{\bar d}_{ \bar X}(\mathbf{k})\nonumber\\
                                 &\times
                                   \left[V_\mathbf{q}O^{a}_{Y}(\mathbf{p+q})O^{b}_{Y}(\mathbf{p})O^{c}_{ \bar Y}(\mathbf{k-q})O^{d}_{ \bar Y}(\mathbf{k})
                                   -V_\mathbf{k-p-q}O^{a}_{ Y}(\mathbf{p+q})O^{b}_{\bar  Y}(\mathbf{p})O^{c}_{ \bar Y}(\mathbf{k-q})O^{d}_{ Y}(\mathbf{k})\right]\,. 
  \end{align}
\end{widetext}

\end{document}